\documentclass[aps,prr,twocolumn,amsmath,amssymb,superscriptaddress,longbibliography]{revtex4-1}
\usepackage[english]{babel}
\usepackage{xcolor}
\usepackage{amssymb}
\usepackage{dcolumn}
\usepackage{bm}
\usepackage{graphicx}
\usepackage{amsmath}
\usepackage{braket}

\usepackage{graphicx}        

\graphicspath{{pict/}{}}

\usepackage[normalem]{ulem}

\usepackage{dcolumn}
\usepackage{bm}
\usepackage[pdfstartview=FitH, CJKbookmarks=true, bookmarksnumbered=true, bookmarksopen=true, colorlinks=true, pdfborder=001, citecolor=blue, linkcolor=blue, urlcolor=blue, linktocpage=true] {hyperref}

\makeatletter

\begin{document}

\title{Kerr-Filtered Hybrid States from Double-Squeezed Tripartite Interactions}

\author{Jing Tang}
\affiliation{School of Physics and Optoelectronic Engineering, Guangdong University of Technology, Guangzhou 510006, China}
\affiliation{Guangdong Provincial Key Laboratory of Sensing Physics and System Integration Applications, Guangdong University of Technology, Guangzhou, 510006, China}

\author{Yuangang Deng}
\email{dengyg3@mail.sysu.edu.cn}
\affiliation{Guangdong Provincial Key Laboratory of Quantum Metrology and Sensing $\&$ School of Physics and Astronomy, Sun Yat-Sen University, Zhuhai 519082, China}

\begin{abstract}
Multipartite interactions provide a unique route to generate hybrid quantum states, yet their intrinsically weak nonlinear coupling limits the controllable manipulation of the resulting correlations. Here we show that cooperative double squeezing and cavity Kerr nonlinearity together enable controllable engineering of Kerr-filtered hybrid states in a tripartite spin-photon-phonon system. Double squeezing exponentially enhances the spin-mediated photon-phonon pair-generation channel, whereas the Kerr interaction selectively reshapes the photonic component while maintaining the underlying tripartite conversion process. This mechanism reconstructs the composition of the hybrid excitation and gives rise to a fourfold phase-space manifold associated with spontaneous $Z_{2}^{(a)}\times Z_{2}^{(b)}$ parity breaking. Our results reveal a strategy for engineering hybrid quantum states by combining interaction amplification with mode-selective nonlinear filtering.
\end{abstract}

\maketitle
\maketitle

{\em Introduction}.---Nonlinear interactions convert quantum fluctuations into correlated quantum states and underpin diverse phenomena in quantum optics, hybrid quantum systems, and quantum information science~\cite{Haroche2013,Aspelmeyer2014,Blais2021,Arute2019}. A qualitatively distinct regime emerges when multiple quantum degrees of freedom are coupled through a single microscopic process~\cite{Clerk2010,Kurizki2015,Lauk2020}. Unlike hybrid architectures based on sequential pairwise interactions~\cite{rabi1936process,jaynes1963comparison,xiang2013hybrid,
shandilya2021optomechanical}, genuine multipartite couplings generate correlations through a shared excitation pathway
~\cite{Hei2023,Tang2024APL,lemonde2016enhanced}, where multiple modes participate in the same quantum event. Consequently, the resulting correlations are encoded not only in the final quantum state but also in the microscopic process responsible for its generation.  This intrinsic connection enables unique opportunities for hybrid entanglement, quantum transduction, and nonclassical-state engineering~\cite{Riedinger2018,Higginbotham2018,Hong1987,Grimm2020,Hei2023}. However, this connection also locks the state composition to the underlying interaction pathway. A central question is therefore whether the composition of a multipartite quantum state can be selectively reconfigured without disrupting the microscopic process that generates its correlations.

Tripartite interactions provide a natural platform for addressing this question, where three quantum degrees of freedom are correlated through a shared microscopic transition.  They enable excitation-conversion and pair-generation processes absent in conventional Rabi or bilinear beam-splitter models~\cite{rabi1936process,jaynes1963comparison,
xiang2013hybrid,shandilya2021optomechanical}, providing new routes toward multipartite entanglement, coherent quantum transduction, and
nonclassical photon-phonon emission~\cite{OckeloenKorppi2018,Chu2018,Forsch2020,Andrews2014}.
In particular, cavity-enhanced Stokes scattering realizes a direct spin-mediated conversion process in which a single atomic excitation is converted into a correlated photon-phonon pair, enabling the observation of vacuum fluctuations and nonclassical single-quanta
emission~\cite{Tang2024APL,Riedinger2016,SafaviNaeini2013,Underwood2015,Hong2017}. Despite these advances, correlation generation and hybrid-state composition remain intrinsically linked. Existing approaches enhance correlation generation but lack an independent mechanism to reshape the hybrid-state composition.

Parametric amplification provides a powerful route to overcoming the weak-coupling limitation by exponentially enhancing effective
interactions~\cite{Liao2019,Qiao2026,Lemonde2016,Qin2018,Wang2022,Hei2023}. This approach has enabled enhanced optomechanical and light-matter couplings~\cite{Didier2015,Lu2015,Lemonde2016,Qin2018,Wang2022}, spin-magnon-mechanical interactions~\cite{Zhang2016,Li2018Magnon,Harder2018,Hei2023,Tang2024APL}, as well as strongly anharmonic squeezed-Fock states from weak Kerr nonlinearities~\cite{Frattini2023,Qiao2026,Cai2026,Meher2026}. In conventional bipartite systems, however, enhancement arises from squeezing a single mode and is accompanied by increased quantum noise~\cite{Kronwald2013,Wollman2015,Lecocq2015,Levitan2016,Metelmann2017,Malz2020}. Tripartite photon-phonon processes offer a distinct opportunity: because photon and phonon excitations are generated through the same microscopic transition~\cite{Tang2024APL,Liu2013PRL,Galland2014PRL,Greve2021Nature,Mayor2025NatCommun}, simultaneous squeezing of both bosonic modes directly amplifies the correlated pair-generation channel. However, double squeezing only enhances correlation generation and leaves the hybrid-state composition unchanged.

In this Letter, we introduce a cavity-Kerr-assisted double-squeezed spin-photon-phonon system. Double squeezing exponentially enhances the spin-mediated photon-phonon pair-generation channel, driving weak tripartite interactions into the collective strong-coupling regime. The cavity Kerr nonlinearity selectively modifies the photon component without disrupting the underlying tripartite conversion channel.  Thus, double squeezing controls correlation generation, whereas Kerr filtering controls hybrid-state composition.

We show that the interplay of these two mechanisms gives rise to a rich hybrid-superradiant phase diagram beyond the conventional pairwise Rabi paradigm~\cite{rabi1936process,jaynes1963comparison}. Besides the normal phase, the system exhibits a weak-superradiant (WSR) phase and a Kerr-filtered hybrid superradiant (SR) phase, where collective ordering emerges with a progressively phonon-dominated composition. Deep in the SR regime, the cavity-phonon state develops a fourfold phase-space structure arising from spontaneous breaking of the generalized $Z_2^{(a)}\times Z_2^{(b)}$ parity symmetry. The transition is further accompanied by excitation-gap closing, enhanced subsystem entropy, and robust nonclassical signatures in the presence of dissipation. These results establish Kerr filtering as a mechanism for controlling the composition and collective order of multipartite quantum states~\cite{Ma2024Nature,Leib2023Nature,Kessler2012PRA,Minganti2020PRA,Kirton2019AdvQuantumTechnol,grimm2020stabilization,he2023fast}.

\begin{figure}[ht]
\includegraphics[width=0.95\columnwidth]{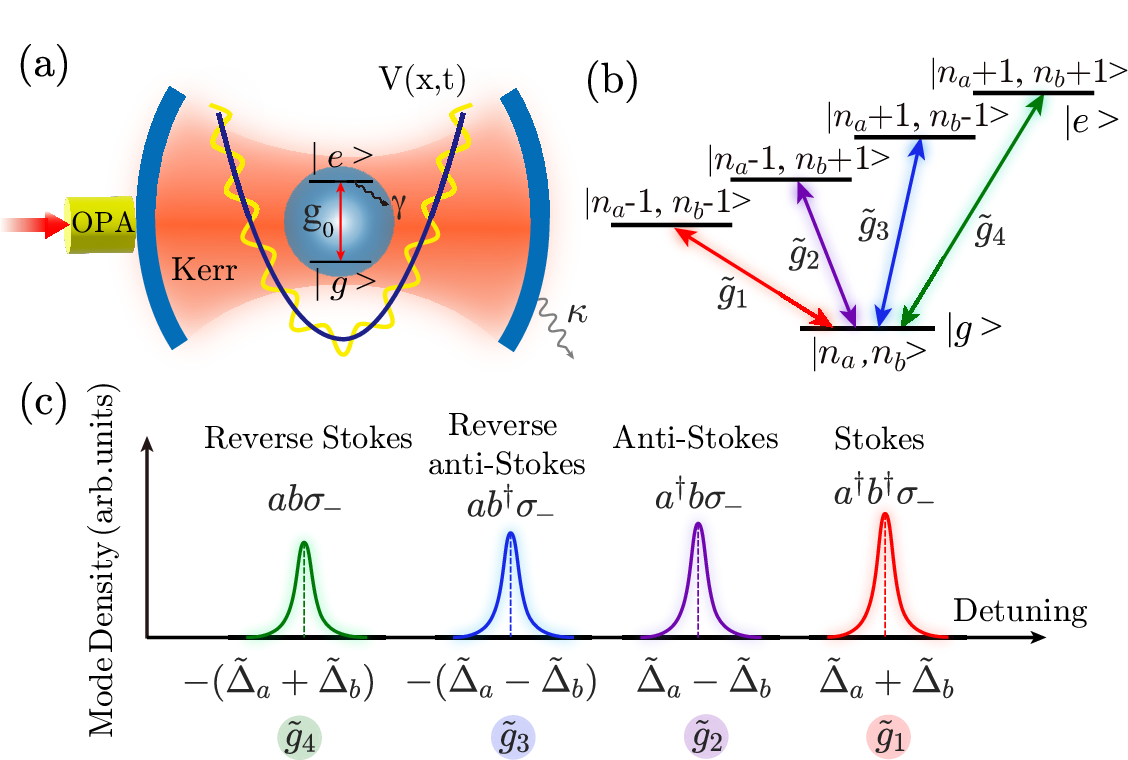}
\caption{(a) Sketch of double-squeezed tripartite spin-photon-phonon interactions. (b) Tripartite beam-splitter and squeezing interactions generated by the double Bogoliubov transformation, corresponding to Stokes, anti-Stokes, and their reverse processes. (c) Scattering picture of the four distinct sideband processes.
}
\label{scheme}
\end{figure}

{\em Model and Hamiltonian}.---As illustrated in Fig.~\ref{scheme}, we consider a single alkaline-earth atom, such as $^{88}$Sr, coupled to a high-finesse optical cavity and confined in a one-dimensional harmonic trap. The cavity mode exhibits a weak Kerr nonlinearity and is parametrically driven with amplitude $\chi_a$, while the trapping potential is modulated near twice the motional frequency, $V(x,t)=  M\omega_b^2[1+\epsilon\cos(2\omega_d t)]\hat x^2/2$. Here $\omega_d$ and $\epsilon$ denote the modulation frequency and depth, respectively. The atom is modeled as a two-level system with ground state $|g\rangle$ ($^1S_0$) and long-lived excited state $|e\rangle$ ($^3P_1$), coupled to the cavity mode with vacuum Rabi frequency $g_0$.

Within Lamb-Dicke regime, the system is described by the tripartite spin-photon-phonon Hamiltonian
\begin{align}
\hat{\cal H}_0/\hbar=&~\Delta_b\hat b^\dagger\hat b+\Delta_a\hat a^\dagger\hat a + \frac{\Delta_s}{2}\sigma_z +g\left(\hat a^\dagger\hat b^\dagger\sigma_- + \hat a\hat b\sigma_+\right) 
\nonumber\\
&+ K_0\hat a^\dagger\hat a^\dagger\hat a\hat a + \frac{\chi_a}{2} \left(\hat a^2+\hat a^{\dagger2}\right) +
\frac{\chi_b}{2}\left( \hat b^2+\hat b^{\dagger2}\right),
\label{H0_pair_full}
\end{align}
where $\hat a$ ($\hat b$) annihilates cavity photon (motional phonon), $\Delta_{a,b,s}$ denote the cavity, phonon, and atomic detunings,
respectively, $\chi_b=\epsilon\omega_b/4$ is the phonon parametric-drive amplitude, and $g=g_0k_Lx_{\rm ZPF}$
with $x_{\rm ZPF}=\sqrt{\hbar/(2M\omega_b)}$. The tripartite term converts an atomic excitation into a correlated photon-phonon pair, defining the microscopic origin of the hybrid correlations studied below.

To reveal the squeezing-enhanced interaction, we diagonalize the bosonic sector using a double Bogoliubov transformation
\begin{equation}
{\cal U}(r_a,r_b)
=
\exp\!\left[
-\frac{r_a}{2}
\left( \hat a^2-\hat a^{\dagger2} \right)
\right]
\exp\!\left[
-\frac{r_b}{2}
\left( \hat b^2-\hat b^{\dagger2} \right)
\right],
\label{double_U_pair_renorm}
\end{equation}
where the squeezing parameters satisfy
$\tanh2r_a=-(\chi_a+\delta\chi_a)/(\Delta_a+\delta\Delta_a)$
and
$\tanh2r_b=-\chi_b/\Delta_b$,
with
$\delta\chi_a=3K_0\sinh4r_a$
and
$\delta\Delta_a=
2K_0\sinh^2r_a(3\cosh2r_a+1)$.
Including the leading Kerr renormalization, the effective Hamiltonian becomes
\begin{align}
\hat{\cal H}_{1}/\hbar=&~\tilde\Delta_b\hat b^\dagger\hat b+\tilde\Delta_a\hat a^\dagger\hat a+\frac{\Delta_s}{2}\sigma_z
+U_a\hat a^{\dagger2}\hat a^2
\nonumber\\
+&
\tilde g_1 (\hat a^\dagger\hat b^\dagger\sigma_-+\hat a\hat b\sigma_+)+\tilde g_2(\hat a^\dagger\hat b\sigma_-+\hat a\hat b^\dagger\sigma_+)
\nonumber\\
+&
\tilde g_3(\hat a\hat b^\dagger\sigma_-+\hat a^\dagger\hat b\sigma_+)+
\tilde g_4(\hat a\hat b\sigma_-+\hat a^\dagger\hat b^\dagger\sigma_+),
\label{Hsq_pair_final}
\end{align}
where
$\tilde g_1=g\cosh r_a\cosh r_b$,
$\tilde g_2=\tilde g_1\tanh r_b$,
$\tilde g_3=\tilde g_1\tanh r_a$,
and
$\tilde g_4=\tilde g_1\tanh r_a\tanh r_b$.
The renormalized detunings are
$\tilde\Delta_a=\sqrt{(\Delta_a+\delta\Delta_a)^2-(\chi_a+\delta\chi_a)^2}$
and
$\tilde\Delta_b=\sqrt{\Delta_b^2-\chi_b^2}$,
while the Kerr interaction is enhanced to
$U_a=[3\cosh(4r_a)+1]K_0/4$.
Equation~(\ref{Hsq_pair_final}) shows that parametric amplification enhances both the tripartite couplings and Kerr nonlinearity, providing the basis for the Kerr-filtered hybrid phases.

Equation~(\ref{Hsq_pair_final}) reveals the microscopic origin of the double-squeezed hybrid dynamics. Unlike the conventional Rabi model with a single bosonic mode coupled to the spin, Eq.~(\ref{Hsq_pair_final}) contains four tripartite scattering channels involving both bosonic modes [Fig.~\ref{scheme}(c)]. Among them, the pair-generation channel $\hat a^\dagger\hat b^\dagger\sigma_-$ is exponentially enhanced, with $\tilde g_1=g\cosh r_a\cosh r_b$. This enhancement arises from the simultaneous squeezing of both modes, which amplifies correlated photon-phonon generation.

Double squeezing also enhances the Kerr nonlinearity more strongly, $U_a\propto e^{4r_a}$, compared with the tripartite interaction, $\tilde g_1\propto e^{r_a+r_b}$. Thus, the tripartite interaction drives correlated excitation buildup, whereas the Kerr nonlinearity filters the photonic component. Their interplay provides the central mechanism of this work, reshaping the hybrid-state composition while preserving its microscopic generation channel.

\begin{figure}[ht]
\includegraphics[width=0.9\columnwidth]{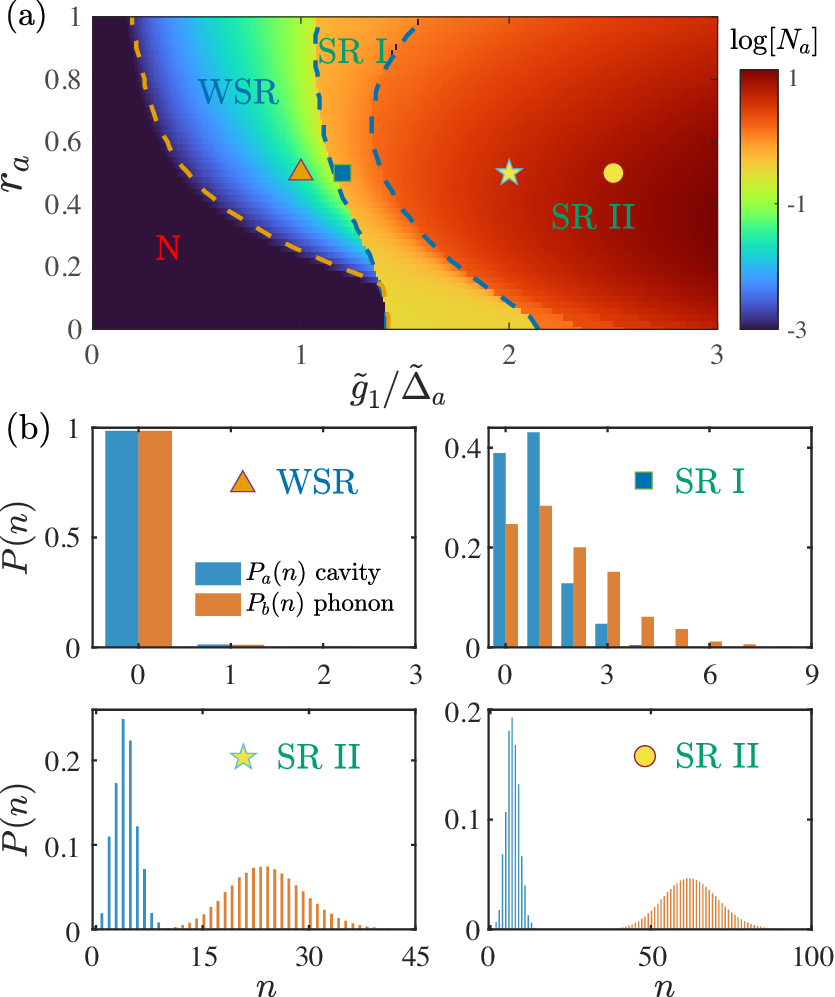}
\caption{Ground-state phase diagram and Kerr-induced reconstruction of the hybrid state. (a) Cavity occupation $N_a=\langle \hat a^\dagger \hat a\rangle$ in $\tilde g_1$-$r_a$ parameter plane. Symbols indicate representative points at $r_a=0.5$ with $\tilde g_1/\tilde\Delta_a=1$, 1.2, 2.0, and 2.5. (b) Corresponding cavity and phonon Fock distributions, $P_c(n)$ and $P_p(n)$, illustrating the Kerr-induced redistribution of the hybrid-state composition. Other parameters are $\tilde\Delta_a=\tilde\Delta_b=\Delta_s=2g$ and $K_0=0.25g$.
}
\label{phase2}
\end{figure}

Equation~(\ref{Hsq_pair_final}) possesses a $Z_{2}^{(a)}\times Z_{2}^{(b)}$ symmetry generated by the generalized parity operators
$\hat{\mathcal P}_{a}=e^{i\pi\hat a^\dagger\hat a}\sigma_z$ and $\hat{\mathcal P}_{b}=e^{i\pi\hat b^\dagger\hat b}\sigma_z$, under which
\begin{align}
\hat{\mathcal P}_{a}:
(\hat a,\hat b,\sigma_-)
&\rightarrow
(-\hat a,\hat b,-\sigma_-),
\nonumber\\
\hat{\mathcal P}_{b}:
(\hat a,\hat b,\sigma_-)
&\rightarrow
(\hat a,-\hat b,-\sigma_-).
\label{ParityTransform}
\end{align}
All four scattering channels are invariant under these independent parity operations. Consequently, spontaneous symmetry breaking produces four symmetry-related ordered states $(\pm\alpha,\pm\beta)$ associated with independent cavity and phonon order parameters. This fourfold manifold defines the symmetry structure of the Kerr-filtered hybrid superradiant phase and is directly visualized by the joint cavity-phonon phase space.

{\em Ground-state phase diagram}.---Figure~\ref{phase2}(a) presents the ground-state phase diagram of the system in $(\tilde g_1,r_a)$ plane for $r_a=r_b$. The phase diagram contains the normal phase (N), a weak-superradiant (WSR) regime, and two hybrid-superradiant regimes (SR I and SR II). The normal phase has vanishing bosonic order parameters, whereas the ordered phases exhibit simultaneous cavity and phonon condensation. Increasing squeezing rapidly lowers the critical coupling through exponential enhancement of the effective tripartite interaction, enabling collective order at weak bare coupling.

\begin{figure}[ht]
\includegraphics[width=0.98\columnwidth]{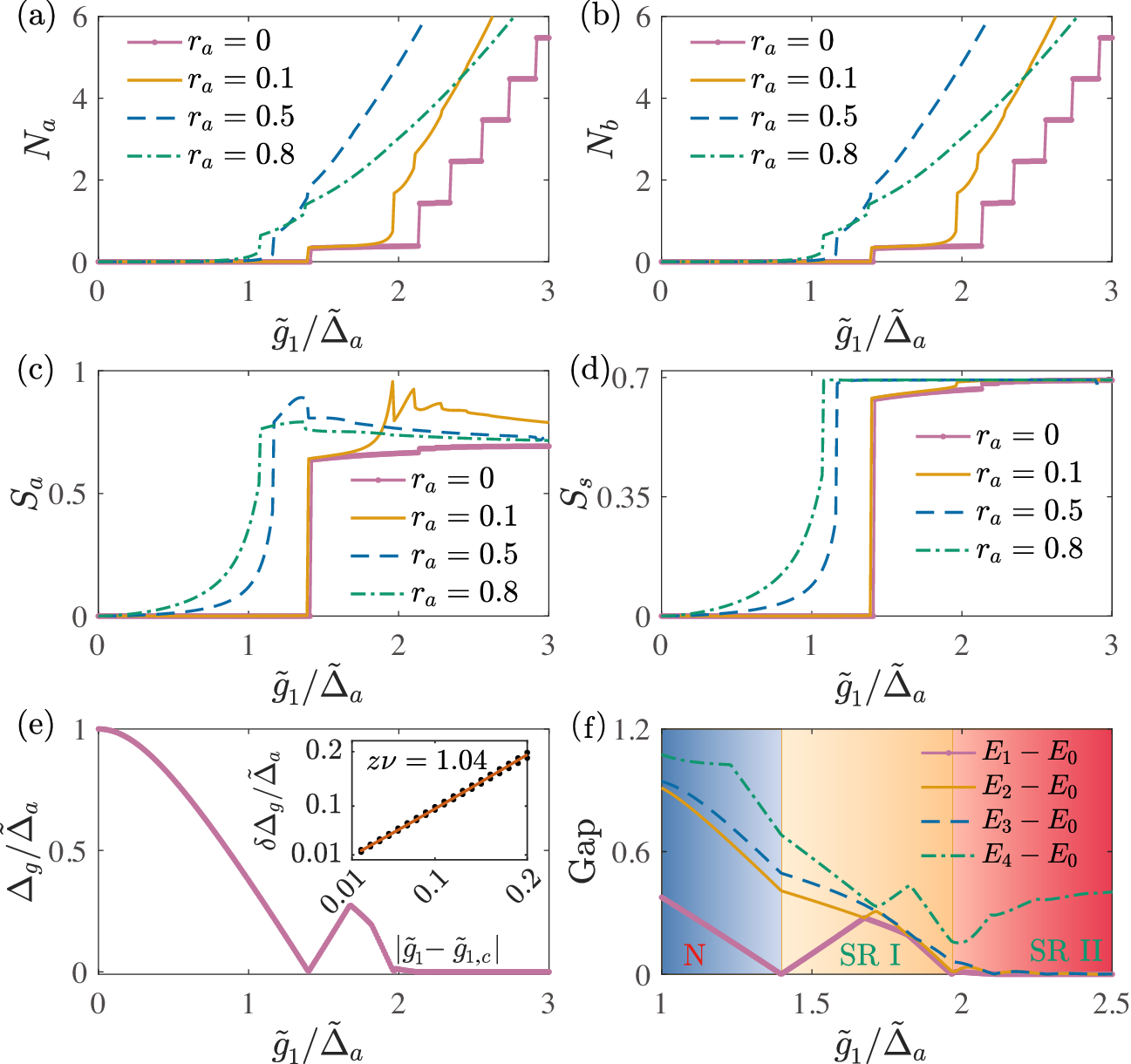}
\caption{Kerr-filtered hybrid superradiance induced by double-squeezed tripartite interactions. (a)-(b) Cavity and phonon occupations $N_a$ and $N_b$, and (c)-(d) cavity and spin entropy $S_a$ and $S_s$ versus $\tilde g_1$ for different $r_a$. (e) Lowest excitation gap $\Delta_g$ for $r_a=0.1$. The inset shows the finite-size scaling of the gap closing, $\delta\Delta_g\propto |\tilde g_1-\tilde g_{1,c}|^{z\nu}$, with $z\nu\simeq1.04$. (f) First four excitation gaps versus $\tilde g_1$ at $r_a=0.1$.
}
\label{orders}
\end{figure}

Although SR I and SR II share the same broken symmetry, they exhibit distinct hybrid-state compositions. In the WSR regime, the state remains close to vacuum with only weak photon-phonon correlations, as shown in Fig.~\ref{phase2}(b). In SR I, both modes acquire finite occupations, while their Fock distributions remain strongly overlapped due to moderate Kerr filtering. Increasing the interaction further induces a qualitative reconstruction: the phonon mode develops a macroscopic occupation, whereas the cavity field remains confined to the
few-photon manifold. Consequently, the cavity and phonon Fock distributions become nearly separated, signaling a transition from a Fock-space-miscible (SR I) to a Fock-space-immiscible (SR II) hybrid superradiant state. This redistribution arises from the different scaling of the tripartite interaction and Kerr nonlinearity: the former drives collective excitation, whereas the latter filters the photonic component while preserving the hybrid superradiant order.

To identify the microscopic origin of the hybrid-superradiant threshold, we consider the weak-squeezing regime where correlated pair generation dominates. The transition is then controlled by the lowest photon-phonon pair dressed manifold, whose lower branch crosses the normal-state energy at $\tilde g_{1,c} =\sqrt{\Delta_s\left(
\tilde\Delta_a+\tilde\Delta_b\right)}$. The Kerr nonlinearity does not directly shift this pair-dominated threshold because the lowest manifold contains at most one cavity photon, for which the Kerr energy vanishes. Its role emerges beyond the onset, when higher excitation manifolds become populated. Minimization of the effective energy landscape yields
\begin{equation}
\frac{N_b}{N_a}
=
\frac{\tilde\Delta_a+2U_aN_a}
{\tilde\Delta_b},
\label{occupation}
\end{equation}
showing that the occupation-dependent Kerr shift progressively suppresses the photonic weight in favor of the phononic component. Thus, double squeezing controls the onset of hybrid ordering through enhanced pair generation, whereas Kerr filtering reconstructs the internal composition of the ordered state.

{\em Emergence of hybrid superradiance}.---We next investigate the formation of the hybrid ordered phase as the effective tripartite interaction increases. Figures~\ref{orders}(a) and \ref{orders}(b) show the cavity and phonon occupations as functions of the effective coupling for different squeezing strengths. Without squeezing ($r_a=0$), both occupations exhibit a sequence of discrete plateaus characteristic of few-body excitation manifolds. Double squeezing qualitatively changes this behavior by exponentially enhancing the tripartite interaction, transforming the staircase excitation into a continuous collective instability and substantially reducing the critical coupling. The system therefore undergoes a continuous transition from the normal phase to a hybrid SR phase with simultaneous cavity and phonon condensation.

Near the critical point, the system enters the WSR regime, where finite order parameters emerge but bosonic occupations remain small. Upon further increasing $\tilde g_1$, the ordered phase evolves smoothly into the strongly SR regime. While the collective instability amplifies correlated photon-phonon excitations, the Kerr nonlinearity selectively filters the photonic component without disrupting the underlying generation channel. Consequently, the ordered phase is progressively reconstructed from a nearly balanced hybrid state into a phonon-dominated SR state, leading to the characteristic hierarchy $N_b\gg N_a$
deep inside the ordered phase.

To characterize the quantum critical behavior, we evaluate the subsystem von Neumann entropies, $S_\mu=-{\rm Tr}(\rho_\mu\ln\rho_\mu)$ and $\rho_\mu={\rm Tr}_{\bar\mu}(\rho)$, where $\mu=a$ and $s$ denote the cavity and spin subsystems, respectively. As shown in Figs.~\ref{orders}(c) and \ref{orders}(d), both entropies increase sharply at the critical coupling where the bosonic order parameters become finite, signaling the rapid buildup of hybrid quantum correlations accompanying the onset of collective order. The entropy plateau in the ordered phase indicates that the Kerr-filtered hybrid state sustains strong quantum correlations over a broad parameter range.

\begin{figure}[ht]
\includegraphics[width=1.0\columnwidth]{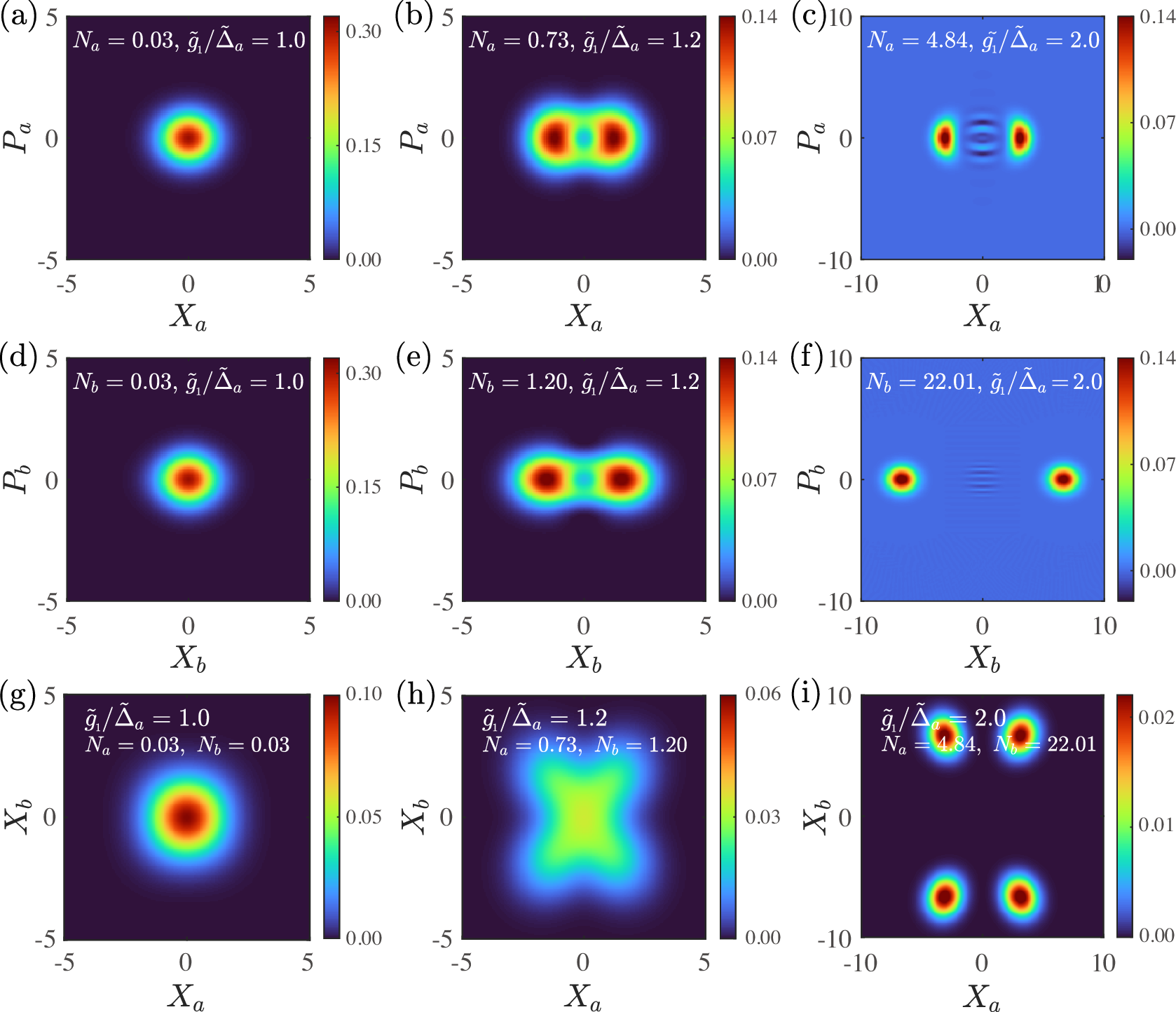}
\caption{Phase-space structure of the hybrid superradiant phases at $r_a=0.5$. Panels (a)-(c) show the cavity Wigner functions $W_a(X_a,P_a)$, and panels (d)-(f) show the phonon Wigner functions $W_b(X_b,P_b)$. Panels (g)-(i) show the corresponding joint cavity-phonon Husimi-$Q$ distributions after tracing out the spin. The three columns correspond to the WSR, SR I, and SR II regimes at $\tilde g_1/\tilde{\Delta}_a=1.0$, 1.2, and 2.0, respectively. The joint Husimi-$Q$ distributions reveal the evolution of the hybrid ordered manifold.}\label{wigner}
\end{figure}

The continuous nature of the transition is further confirmed by the excitation spectrum. Figure~\ref{orders}(e) shows that the excitation gap
$\Delta_g=E_1-E_0$ closes continuously at the critical point and reopens inside the ordered phase. The inset exhibits the finite-size scaling
$\Delta_g\propto|\tilde g_1-\tilde g_{1,c}|^{z\nu}$, yielding $z\nu\simeq1.04$. This critical exponent differs markedly from the Rabi universality class
($z\nu=1/2$)~\cite{PhysRevLett.117.123602,PhysRevLett.115.180404}, indicating that the transition is governed by correlated tripartite photon--phonon fluctuations rather than single-mode criticality.

The higher excitation spectrum provides a direct signature of the underlying symmetry breaking. As shown in Fig.~\ref{orders}(f), the lowest four excitation gaps collapse simultaneously upon entering the ordered phase, forming a fourfold-degenerate low-energy manifold.
This fourfold structure originates from the $Z_2^{(a)}\times Z_2^{(b)}$ symmetry of Eq.~(\ref{Hsq_pair_final}), whose spontaneous breaking yields four symmetry-related states $(\pm\alpha,\pm\beta)$. Together with finite bosonic order and enhanced subsystem entropy, these spectral features establish the Kerr-filtered hybrid superradiant phase with fourfold symmetry-breaking order.

Although SR I and SR II are associated with the same underlying $Z_2^{(a)}\times Z_2^{(b)}$ symmetry, they are distinguished by a qualitative reconstruction of the low-energy manifold. The SR I to SR II boundary is accompanied by the closing and reopening of the excitation gap and a change in the ground-state degeneracy. These spectral and phase-space signatures identify SR I and SR II as distinct hybrid superradiant phases rather than a smooth crossover.

{\em Phase-space structure of hybrid superradiant phases}.---To reveal the phase-space structure of the hybrid ordered phases, we examine the reduced Wigner functions of the cavity and phonon modes together with the joint cavity-phonon Husimi-$Q$ distribution, as shown
in Fig.~\ref{wigner}. The reduced Wigner functions characterize the nonclassical properties of the individual bosonic modes, whereas the
joint Husimi-$Q$ distribution directly reveals the symmetry of the hybrid ordered manifold.

\begin{figure}[ht]
\includegraphics[width=0.95\columnwidth]{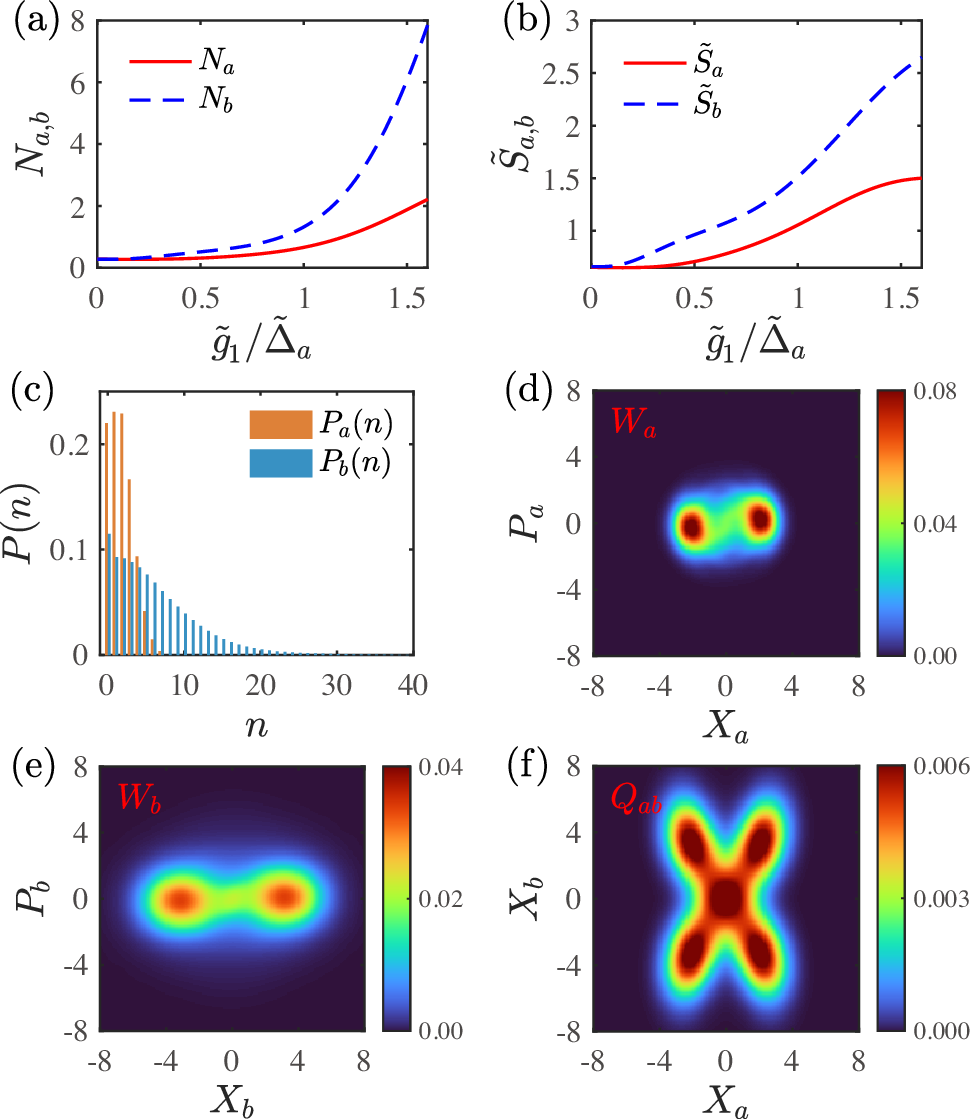}
\caption{Dissipative signatures of the Kerr-filtered hybrid superradiant phase at $r_a=0.5$. (a) Steady-state cavity and phonon occupations $N_{a,b}$ and (b) reduced von Neumann entropies $\tilde S_{a,b}$ versus $\tilde g_1/\tilde\Delta_a$. (c)-(f) Steady-state quantum-state characterization at $\tilde{g}_1/\tilde{\Delta}_a=1.5$: (c) typical cavity and phonon Fock-state distributions $P_a(n)$ and $P_b(n)$, (d) cavity Wigner function $W_a$, (e) phonon Wigner function $W_b$, and (f) joint cavity-phonon Husimi-$Q$ distribution $Q_{ab}$ after tracing out the spin.
}\label{SS_wigner}
\end{figure}

In WSR regime [Figs.~\ref{wigner}(a), \ref{wigner}(d), and \ref{wigner}(g)], both reduced Wigner functions remain nearly Gaussian and localized around the vacuum, while the joint Husimi-$Q$ distribution exhibits a single maximum centered at the origin. Entering the SR I phase
[Figs.~\ref{wigner}(b), \ref{wigner}(e), and \ref{wigner}(h)], the reduced Wigner functions split into two symmetry-related lobes, signaling the emergence of macroscopic bosonic order. The corresponding joint Husimi-$Q$ distribution, however, remains connected, indicating that the cavity and phonon order parameters still occupy a single collective basin despite the onset of symmetry breaking.

A distinct phase-space structure emerges in the SR II phase. As shown in Figs.~\ref{wigner}(c), \ref{wigner}(f), and \ref{wigner}(i), both reduced Wigner functions develop pronounced interference fringes and negative regions, demonstrating strong nonclassical coherence beyond a coherent-state description. More importantly, the joint Husimi-$Q$ distribution splits into four maxima at $(\pm\alpha,\pm\beta)$, directly revealing the four symmetry-related configurations of the $Z_2^{(a)}\times Z_2^{(b)}$ manifold of Eq.~(\ref{Hsq_pair_final}). The joint Husimi-$Q$ distribution further resolves this spectral reconstruction in phase space, evolving into four well-separated symmetry-related maxima in SR II. These four peaks provide a phase-space manifestation of the nearly degenerate low-energy manifold, resolving its four symmetry-related configurations without implying individually symmetry-broken eigenstates. Thus, the phase-space evolution directly visualizes the Kerr-filtered reconstruction of the hybrid ordered manifold while preserving the underlying tripartite collective order. These results establish the fourfold phase-space structure of the Kerr-filtered hybrid superradiant phase. We next ask whether this ordered manifold remains robust against dissipation.

{\em Dissipative hybrid superradiance}.---To address this question, we consider the open-system dynamics governed by the master equation
\begin{align}
\frac{d\rho}{dt}
&=
-i[\hat{\cal H}_{1},\rho] + \kappa_a(N_a+1)\mathcal D[\hat a]\rho + \kappa_a N_a\mathcal D[\hat a^\dagger]\rho
\nonumber\\
&+\kappa_a M_a\mathcal S[\hat a]\rho+\kappa_a M_a\mathcal S[\hat a^\dagger]\rho+\kappa_b(N_b+1)\mathcal D[\hat b]\rho
\nonumber\\
&+\kappa_b N_b\mathcal D[\hat b^\dagger]\rho
+
\kappa_b M_b\mathcal S[\hat b]\rho
+
\kappa_b M_b\mathcal S[\hat b^\dagger]\rho
\nonumber\\
&+
\gamma\mathcal D[\sigma_-]\rho,
\label{ME_double_squeezed_final}
\end{align}
where $\kappa_a$, $\kappa_b$, and $\gamma$ denote the cavity, phonon, and spin decay rates, respectively. The Bogoliubov transformation maps the physical vacuum reservoirs onto effective squeezed reservoirs characterized by
$N_\mu=\sinh^2r_\mu$
and
$M_\mu=\cosh r_\mu\sinh r_\mu$
($\mu=a,b$).
Here
$\mathcal D[\hat o]\rho
=\hat o\rho\hat o^\dagger
-(\hat o^\dagger\hat o\rho+\rho\hat o^\dagger\hat o)/2$
is the standard Lindblad dissipator, whereas
$\mathcal S[\hat o]\rho
=\hat o\rho\hat o
-(\hat o^2\rho+\rho\hat o^2)/2$
describes phase-sensitive two-quasiparticle dissipation.

The squeezed reservoirs introduce normal and anomalous dissipation channels. The terms proportional to $\mathcal D[\hat a^\dagger]$
and $\mathcal D[\hat b^\dagger]$ represent effective heating of the Bogoliubov quasiparticles, while the anomalous dissipators
$\mathcal S[\hat a]$, $\mathcal S[\hat a^\dagger]$, $\mathcal S[\hat b]$, and $\mathcal S[\hat b^\dagger]$ encode phase-sensitive pair fluctuations inherited from the squeezed reservoirs. As shown below, despite these additional quantum-noise channels, the Kerr-filtered hybrid superradiant phase remains robust and retains its essential symmetry and correlation properties in the steady state.

Figure~\ref{SS_wigner} summarizes the steady-state properties of the open system. As shown in Fig.~\ref{SS_wigner}(a), both the cavity and phonon occupations increase continuously with the effective tripartite interaction, confirming the persistence of hybrid superradiant ordering under dissipation. Consistent with the equilibrium results, the phonon population grows much faster than the cavity
population, confirming that Kerr filtering redistributes the hybrid composition while preserving collective ordering.

The steady-state hybrid order is accompanied by a rapid increase of the reduced von Neumann entropies,
\begin{equation}
\tilde S_\mu=
-{\rm Tr}
\left(
\tilde\rho_\mu
\ln\tilde\rho_\mu
\right),
\qquad
\mu=a,b,
\end{equation}
where $\tilde\rho_\mu$ denotes the reduced steady-state density matrix of the cavity or phonon subsystem. Unlike the ground-state entropy, $\tilde S_\mu$ quantifies steady-state mixedness arising from coherent correlations and dissipation. As shown in Fig.~\ref{SS_wigner}(b), the phonon entropy increases much more rapidly than the cavity entropy, consistent with the strongly enhanced phonon occupation and the Kerr-filtered redistribution of the hybrid ordered state.

Further insight is provided by the steady-state phase-space distributions. The broad Fock-state distributions [Fig.~\ref{SS_wigner}(c)] demonstrate that the dissipative steady state extends over many excitation manifolds, while the reduced Wigner functions [Figs.~\ref{SS_wigner}(d) and \ref{SS_wigner}(e)] develop well-separated double-lobed structures characteristic of macroscopic bosonic order.
More importantly, the joint cavity-phonon Husimi-$Q$ distribution [Fig.~\ref{SS_wigner}(f)] forms four symmetry-related maxima located at
$(\pm\alpha,\pm\beta)$, faithfully reproducing the $Z_2^{(a)}\times Z_2^{(b)}$ ordered manifold established in the closed system. These results demonstrate that collective order, fourfold symmetry, and phase-space structure of the Kerr-filtered hybrid phase survive dissipation as a genuine nonequilibrium steady state.

{\em Conclusion}.---In summary, we have shown that double squeezing and cavity Kerr nonlinearity provide two distinct controls over multipartite quantum states. Double squeezing exponentially enhances the spin-mediated photon-phonon pair-generation channel and drives hybrid superradiant ordering, whereas Kerr filtering selectively reshapes the hybrid-state composition while preserving the underlying tripartite generation channel. This separation between correlation generation and state composition control gives rise to hybrid superradiant phases beyond conventional Rabi paradigms. Deep in the ordered regime, the system develops a fourfold phase-space manifold from spontaneous $Z_{2}^{(a)}\times Z_{2}^{(b)}$ symmetry breaking, together with gap closing, enhanced subsystem entropy, and robust dissipative nonclassical signatures.

More broadly, our work identifies Kerr filtering as a route for controlling the composition of multipartite states while preserving their microscopic generation process. The combination of parametric amplification and mode-selective nonlinear filtering provides a strategy for engineering hybrid quantum states with controllable composition and collective order~\cite{Frattini2024Nature,Dutt2024,GonzalezTudela2024}. Since the required ingredients have been demonstrated in cavity-QED~\cite{Ritsch2013,Baumann2010,Norcia2018}, optomechanical~\cite{Aspelmeyer2014}, superconducting-circuit~\cite{Blais2021}, and spin-magnon hybrid systems~\cite{Huebl2013,Tabuchi2015}, our results open a broadly applicable route toward realizing nonequilibrium hybrid quantum phases
~\cite{Ritsch2013,Mivehvar2021}, programmable symmetry breaking ~\cite{Baumann2010,Dogra2019Science},
and multipartite quantum-state engineering through nonlinear interaction engineering
~\cite{Qin2018,Frattini2023}.

{\em Acknowledgments}.---This work was supported by the National Natural Science Foundation of China (Grant No. 12135018, Grant No. 12274473, and Grant No.12374365), Quantum Science and Technology-National Science and Technology Major Project (Grant No.2025ZD0300400), Guangdong Provincial Quantum Science Strategic Initiative (Grants No. GDZX2505001 and No. GDZX2604004), and Guangdong University of Technology SPOE Seed Foundation (SF2024111504).


\twocolumngrid

\appendix

\section{Derivation of the tripartite Hamiltonian\label{sm_model}}

In this section, we derive the tripartite spin-photon-phonon Hamiltonian  starting from the cavity-coupled atomic system illustrated in Fig.~1 of the main text. We consider a single alkaline-earth atom confined in a single-mode high-finesse cavity with bare frequency $\omega_c$. For concreteness, we consider the two relevant electronic states of $^{88}$Sr: the ground state $|g\rangle(^1S_0)$ and the long-lived excited state $|e\rangle(^3P_1)$. The narrow dipole-forbidden transition $|g\rangle\leftrightarrow|e\rangle$ has linewidth $7.5$ kHz, corresponding to wavelength $\lambda=689$ nm and wave vector $k_L=2\pi/\lambda$.

The atom couples to the cavity mode through the spatially dependent interaction $g_0\sin(k_Lx)$, where $g_0$ denotes the maximum single-atom coupling strength. The cavity mode further includes a weak Kerr nonlinearity and a parametric drive with amplitude $\chi_a$. The Hamiltonian before transforming to the rotating frame is
\begin{align}
\hat{\cal H}_0/\hbar
={}&
\frac{\hat p_x^2}{2M}
+V(x,t)
+\frac{\Delta}{2}\sigma_z
+\Delta_a\hat a^\dagger\hat a
\nonumber\\
&+
g_0\sin(k_Lx)
\left(
\hat a^\dagger\hat\sigma_-
+\hat a\hat\sigma_+
\right)
\nonumber\\
&+
K_0\hat a^\dagger\hat a^\dagger\hat a\hat a
+\frac{\chi_a}{2}
\left(
\hat a^2+\hat a^{\dagger2}
\right),
\label{SM-single}
\end{align}
where $M$ is the atomic mass, $\hat a$ is the cavity annihilation operator, and $\sigma_{\pm}$ denote the atomic raising and lowering operators. The detunings $\Delta_a$ and $\Delta$ represent the cavity and atomic detunings, respectively.

The atomic motion is confined along one dimension by a parametrically driven harmonic trap,
\[
V(x,t)=\frac12M\omega_b^2[1+\epsilon\cos(2\omega_dt)]\hat x^2 .
\]
Here $\omega_b$ is the bare trap frequency, while $\omega_d$ and $\epsilon$ denote the modulation frequency and amplitude, respectively. The parametric drive is applied near the second harmonic of the motional frequency, enabling phonon-pair creation. The corresponding motional Hamiltonian is
\begin{equation}
\hat {\cal H}_b(t)
=
\frac{\hat p_x^2}{2M}
+
\frac12 M\omega_b^2
\left[
1+\epsilon\cos(2\omega_d t)
\right]\hat x^2 ,
\label{Hb_time}
\end{equation}
Introducing the zero-point fluctuation amplitude
\[
x_{\rm zpf}=\sqrt{\frac{\hbar}{2M\omega_b}},
\]
the position operator is written as
\[
\hat x=x_{\rm zpf}(\hat b+\hat b^\dagger).
\]
Then unmodulated part then becomes
\begin{equation}
\hat {\cal H}_{b,0}
=
\hbar\omega_b
\left(
\hat b^\dagger\hat b+\frac12
\right),
\end{equation}
The modulation-induced term is
\begin{align}
\hat {\cal H}_{\rm mod}(t)
={}&
\frac{1}{2}M\epsilon\omega_b^2
\cos(2\omega_d t)
x_{\rm zpf}^2
(\hat b+\hat b^\dagger)^2
\nonumber\\
={}&
\frac{\hbar\epsilon\omega_b}{4}
\cos(2\omega_d t)
\left(
\hat b^2+\hat b^{\dagger2}
+2\hat b^\dagger\hat b+1
\right).
\end{align}
The terms $\hat b^2$ and $\hat b^{\dagger2}$ correspond to parametric creation and annihilation of phonon pairs.

Moving to the rotating frame at $\omega_d$ and applying the rotating-wave approximation near parametric resonance ($\omega_d\simeq\omega_b$), we obtain
\begin{equation}
\hat {\cal H}_b^{\rm RWA}/\hbar
=
\Delta_b \hat b^\dagger\hat b
+
\frac{\chi_b}{2}
\left(
\hat b^2+\hat b^{\dagger 2}
\right),
\label{Hb_RWA}
\end{equation}
with $\Delta_b=\omega_b-\omega_d$ and $\chi_b={\epsilon\omega_b}/{4}$. The second term describes coherent phonon squeezing induced by the parametric modulation.

In the Lamb-Dicke regime, $k_Lx_{\rm zpf}\ll1$, the cavity coupling can be expanded to first order as $\sin(k_Lx)\simeq k_Lx$. The atom-cavity interaction therefore becomes
\begin{align}
\hat{\cal H}_2/\hbar
={}&
\hat{\cal H}_b^{\rm RWA}/\hbar
+\Delta_a\hat a^\dagger\hat a
+\frac{\Delta}{2}\sigma_z
\nonumber\\
&+
g
\left(
\hat b^\dagger e^{i\omega_dt}
+\hat b e^{-i\omega_dt}
\right)
\left(
\hat a^\dagger\hat\sigma_-
+\hat a\hat\sigma_+
\right)
\nonumber\\
&+
K_0\hat a^\dagger\hat a^\dagger\hat a\hat a
+\frac{\chi_a}{2}
\left(
\hat a^2+\hat a^{\dagger2}
\right),
\label{Ham}
\end{align}

To remove the explicit time dependence associated with the atomic transition, we further transform into the interaction picture defined by ${\cal U}=e^{-i\omega_d\sigma^+\sigma^-t}$. Keeping only the resonant terms yields the effective spin-mediated photon-phonon pair-generation interaction, then Eq.(\ref{Ham}) reduces to
\begin{align}
\hat{\cal H}_3/\hbar
={}&
\hat{\cal H}_b^{\rm RWA}/\hbar
+\Delta_a\hat a^\dagger\hat a
+\frac{\Delta_s}{2}\sigma_z
\nonumber\\
&+
g
\left(
\hat b^\dagger e^{i\omega_dt}
+\hat b e^{-i\omega_dt}
\right)
\left(
\hat a^\dagger\hat\sigma_-e^{-i\omega_dt}
+\hat a\hat\sigma_+e^{i\omega_dt}
\right)
\nonumber\\
&+
K_0\hat a^\dagger\hat a^\dagger\hat a\hat a
+\frac{\chi_a}{2}
\left(
\hat a^2+\hat a^{\dagger2}
\right)
\nonumber\\
\approx{}&
\Delta_b\hat b^\dagger\hat b
+\Delta_a\hat a^\dagger\hat a
+\frac{\Delta_s}{2}\sigma_z
\nonumber\\
&+
g
\left(
\hat a^\dagger\hat b^\dagger\hat\sigma_-
+\hat a\hat b\hat\sigma_+
\right)
+K_0\hat a^\dagger\hat a^\dagger\hat a\hat a
\nonumber\\
&+
\frac{\chi_a}{2}
\left(
\hat a^2+\hat a^{\dagger2}
\right)
+
\frac{\chi_b}{2}
\left(
\hat b^2+\hat b^{\dagger2}
\right),
\label{Ham1}
\end{align}
with the effective atom-light detuning $\Delta_s=\Delta-\omega_d$.

To diagonalize the quadratic bosonic sectors and incorporate the leading Kerr-induced renormalization, we apply the double-squeezing transformation
\begin{align}
{\cal U}(r_a,r_b)
={}&
\exp\!\left[
-\frac{r_a}{2}
\left(
\hat a^2-\hat a^{\dagger2}
\right)
\right]
\nonumber\\
&\times
\exp\!\left[
-\frac{r_b}{2}
\left(
\hat b^2-\hat b^{\dagger2}
\right)
\right].
\label{double_U_pair_renorm}
\end{align}
The transformed operators are
\begin{align}
{\cal U}^\dagger \hat a{\cal U}
&=
\hat a\cosh r_a+\hat a^\dagger\sinh r_a,
\nonumber\\
{\cal U}^\dagger \hat b{\cal U}
&=
\hat b\cosh r_b+\hat b^\dagger\sinh r_b .
\end{align}

The Kerr term $\hat{H}_K = K_0 \hat{a}^{\dagger 2} \hat{a}^2$ requires special treatment because the Bogoliubov transformation generates additional quadratic contributions. The Kerr nonlinearity therefore modifies not only the quartic interaction strength but also the effective quadratic squeezing dynamics. Under the Bogoliubov transformation, $\hat{a} \to \hat{a} \cosh r_a + \hat{a}^\dagger \sinh r_a$, the quartic Kerr interaction generates both quartic and quadratic terms after normal ordering.

Keeping the leading quadratic corrections, we obtain
\begin{align}
{\cal U}^\dagger
K_0\hat a^{\dagger2}\hat a^2
{\cal U}
\simeq{}&
U_a\hat a^{\dagger2}\hat a^2
+\delta\Delta_{a,\text{Kerr}}
\hat a^\dagger\hat a
\nonumber\\
&+
\frac{\delta\chi_{a,\text{Kerr}}}{2}
\left(
\hat a^2+\hat a^{\dagger2}
\right)
+E_K,
\end{align}
where the non-resonant quadratic terms ${\delta\chi_{a, \text{Kerr}}} (\hat{a}^2 + \hat{a}^{\dagger 2})/2$ represent the Kerr-induced pairing feedback. To eliminate the residual quadratic pairing terms, the squeezing parameter is determined self-consistently by
\begin{equation}
\tanh 2r_a
=
-
\frac{
\chi_a+\delta\chi_{a,\text{Kerr}}
}{
\Delta_a+\delta\Delta_{a,\text{Kerr}}
}.
\end{equation}
The renormalized parameters are given by:
\begin{align}
U_a
={}&
K_0
\left(
\cosh^4r_a
+4\cosh^2r_a\sinh^2r_a
+\sinh^4r_a
\right)
\nonumber\\
={}&
\frac{3\cosh(4r_a)+1}{4}K_0 ,
\\
\delta\Delta_{a,\rm Kerr}
={}&
K_0
\left(
8\cosh^2r_a\sinh^2r_a
+4\sinh^4r_a
\right)
\nonumber\\
={}&
2K_0\sinh^2r_a
\left(
3\cosh 2r_a+1
\right),
\\
\delta\chi_{a,\rm Kerr}
={}&
6K_0\cosh r_a\sinh r_a\cosh 2r_a
\nonumber\\
={}&
3K_0\sinh 4r_a .
\end{align}
The cavity squeezing parameter is therefore determined self-consistently as
\begin{equation}
\tanh 2r_a
=
-
\frac{
\chi_a+\delta\chi_{a,\rm Kerr}
}{
\Delta_a+\delta\Delta_{a,\rm Kerr}
}.
\label{rab_condition_self_consistent}
\end{equation}
These expressions show that the Kerr nonlinearity is exponentially enhanced under squeezing and simultaneously renormalizes the effective cavity detuning and parametric drive. For the phonon mode, no Kerr correction is present, and the squeezing parameter remains $
\tanh 2r_b=-{\chi_b}/{\Delta_b}$. The renormalized detunings are strictly determined via the self-consistent Bogoliubov forms
\begin{align}
\tilde\Delta_a
={}&
\sqrt{
(\Delta_a+\delta\Delta_{a,\rm Kerr})^2
-
(\chi_a+\delta\chi_{a,\rm Kerr})^2
},
\nonumber\\
\tilde\Delta_b
={}&
\sqrt{\Delta_b^2-\chi_b^2}.
\end{align}

The tripartite interaction transforms into four scattering channels,
\begin{align}
&{\cal U}^\dagger
g\left(
\hat a^\dagger\hat b^\dagger\sigma_-
+\hat a\hat b\sigma_+
\right)
{\cal U}
\nonumber\\
={}&
\tilde g_1
\left(
\hat a^\dagger\hat b^\dagger\sigma_-
+\hat a\hat b\sigma_+
\right)
\nonumber\\
&+
\tilde g_2
\left(
\hat a^\dagger\hat b\sigma_-
+\hat a\hat b^\dagger\sigma_+
\right)
\nonumber\\
&+
\tilde g_3
\left(
\hat a\hat b^\dagger\sigma_-
+\hat a^\dagger\hat b\sigma_+
\right)
\nonumber\\
&+
\tilde g_4
\left(
\hat a\hat b\sigma_-
+\hat a^\dagger\hat b^\dagger\sigma_+
\right),
\label{Hint_transformed}
\end{align}
where the transformed coupling strengths are defined as $\tilde g_1=g\cosh r_a\cosh r_b$,
$\tilde g_2=\tilde g_1\tanh r_b$,
$\tilde g_3=\tilde g_1\tanh r_a$,
and
$\tilde g_4=\tilde g_1\tanh r_a\tanh r_b$. The first term corresponds to the original pair-generation process, whereas the remaining channels arise from the squeezed quadratures.

Omitting irrelevant constant energy shifts, the effective Hamiltonian within
this leading-order Kerr self-consistent approximation is simplified to
\begin{align}
\hat{\cal H}_{4}/\hbar
={}&
\tilde\Delta_b\hat b^\dagger\hat b
+\tilde\Delta_a\hat a^\dagger\hat a
+\frac{\Delta_s}{2}\sigma_z
+U_a\hat a^{\dagger2}\hat a^2
\nonumber\\
&+
\tilde g_1
\left(
\hat a^\dagger\hat b^\dagger\sigma_-
+\hat a\hat b\sigma_+
\right)
\nonumber\\
&+
\tilde g_2
\left(
\hat a^\dagger\hat b\sigma_-
+\hat a\hat b^\dagger\sigma_+
\right)
\nonumber\\
&+
\tilde g_3
\left(
\hat a\hat b^\dagger\sigma_-
+\hat a^\dagger\hat b\sigma_+
\right)
\nonumber\\
&+
\tilde g_4
\left(
\hat a\hat b\sigma_-
+\hat a^\dagger\hat b^\dagger\sigma_+
\right).
\label{Hsq_pair_final}
\end{align}
This Hamiltonian captures the competition between squeezing-enhanced tripartite correlations and Kerr-induced photon-selective nonlinear filtering that governs the hybrid-state reconstruction discussed in the main text.

\section{Analytical origin of Kerr-filtered superradiance}
\label{app:kerr_origin}

We now clarify the distinct microscopic roles of the squeezing-enhanced tripartite interaction and the cavity Kerr nonlinearity. In the weak-squeezing regime, the Bogoliubov-induced scattering is dominated by the correlated pair-generation channel. The corresponding reduced Hamiltonian reads
\begin{align}
\hat{\cal H}_{\rm pair}/\hbar
={}&
\tilde\Delta_a\hat a^\dagger\hat a
+\tilde\Delta_b\hat b^\dagger\hat b
+\frac{\Delta_s}{2}\sigma_z
+U_a\hat a^{\dagger 2}\hat a^2
\nonumber\\
&+
\tilde g_1
\left(
\hat a^\dagger\hat b^\dagger\sigma_-
+\hat a\hat b\sigma_+
\right),
\label{H_pair_app}
\end{align}
This form separates two physically distinct ingredients: the tripartite interaction coherently converts a spin excitation into a correlated photon-phonon pair, whereas the Kerr term penalizes multiple cavity photons and therefore becomes increasingly important at larger excitation number.

We first determine the onset of the hybrid ordered state. For the parameter regime considered in the main text, $\Delta_s>0$, the normal state is
\begin{equation}
|{\rm N}\rangle=|0_a,0_b,g\rangle,
\qquad
E_{\rm N}/\hbar=-\frac{\Delta_s}{2},
\label{normal_energy_app}
\end{equation}
The lowest excitation generated by the tripartite interaction belongs to the correlated pair manifold spanned by
$|0_a,0_b,e\rangle$ and $|1_a,1_b,g\rangle$. Projecting Eq.~(\ref{H_pair_app}) onto this subspace gives
\begin{equation}
\hat{\cal H}_{1}/\hbar=
\begin{pmatrix}
\Delta_s/2
&
\tilde g_1
\\[2mm]
\tilde g_1
&
\tilde\Delta_a+\tilde\Delta_b-\Delta_s/2
\end{pmatrix},
\label{pair_matrix}
\end{equation}
A key point is that the Kerr nonlinearity is inactive in this lowest pair manifold because $\hat a^{\dagger2}\hat a^2|1_a\rangle=0$. The onset of the instability is therefore governed entirely by the hybridization between the atomic excitation and the lowest correlated photon-phonon pair.

Diagonalizing Eq.~(\ref{pair_matrix}) gives the two dressed branches
\begin{equation}
{E_{\pm}}/{\hbar}
=
\frac{\tilde\Delta_a+\tilde\Delta_b}{2}
\pm
\frac{1}{2}
\sqrt{
\left(
\Delta_s-\tilde\Delta_a-\tilde\Delta_b
\right)^2
+4\tilde g_1^2
},
\label{pair_dressed_energy}
\end{equation}
Within the pair-dominated description, the hybrid-superradiant threshold is reached when the lower dressed branch becomes degenerate with the normal state,
\begin{equation}
E_-=E_{\rm N}.
\label{crossing_condition}
\end{equation}
Using Eqs.~(\ref{normal_energy_app}) and
(\ref{pair_dressed_energy}), this condition yields
\begin{align}
\tilde g_{1,c}^2
=
\Delta_s
\left(
\tilde\Delta_a+\tilde\Delta_b
\right), 
\label{critical_general}
\end{align}
or
\begin{equation}
\tilde g_{1,c}
=
\sqrt{
\Delta_s
\left(
\tilde\Delta_a+\tilde\Delta_b
\right)
}.
\label{critical_general_root}
\end{equation}
Equation~(\ref{critical_general_root}) provides a simple spectral interpretation of the transition: hybrid ordering sets in when the energy lowering produced by coherent spin-pair hybridization compensates the excitation cost of the lowest correlated photon-phonon manifold. The absence of $U_a$ at this level is not a consequence of neglecting the Kerr nonlinearity, but follows directly from the single-photon character of the lowest pair excitation. Consequently, Kerr filtering is weak at the onset but becomes progressively important once higher photon manifolds are populated.

We next examine how the Kerr interaction modifies the internal
photon--phonon composition beyond the onset of hybrid ordering. Introducing the semiclassical amplitudes
$\alpha=\langle\hat a\rangle$,
$\beta=\langle\hat b\rangle$,
$s=\langle\sigma_-\rangle$, and
$z=\langle\sigma_z\rangle$, Eq.~(\ref{H_pair_app}) gives the
mean-field energy
\begin{align}
{E_{\rm MF}}/{\hbar}
={}&
\tilde\Delta_a|\alpha|^2
+\tilde\Delta_b|\beta|^2
+U_a|\alpha|^4
+\frac{\Delta_s}{2}z
\nonumber\\
&+
\tilde g_1
\left(
\alpha^\ast\beta^\ast s
+\alpha\beta s^\ast
\right),
\label{EMF_app}
\end{align}
For a pure two-level state, the spin amplitudes satisfy
\begin{equation}
z^2+4|s|^2=1.
\label{spin_constraint_app}
\end{equation}
Minimizing over the spin degree of freedom and defining $N_a=|\alpha|^2$ and $N_b=|\beta|^2$ gives the lower adiabatic energy surface
\begin{equation}
{E_g}/{\hbar}
=
\tilde\Delta_aN_a
+\tilde\Delta_bN_b
+U_aN_a^2
-\frac{1}{2}
\sqrt{
\Delta_s^2
+4\tilde g_1^2N_aN_b
}.
\label{Eg_app}
\end{equation}
The first two terms describe the bare photon and phonon excitation costs, while the square-root contribution represents the energy gained from coherent spin-mediated pair conversion. In contrast, the Kerr term introduces an occupation-dependent energy penalty exclusively in the photonic sector.

\begin{figure}[ht]
\includegraphics[width=0.97\columnwidth]{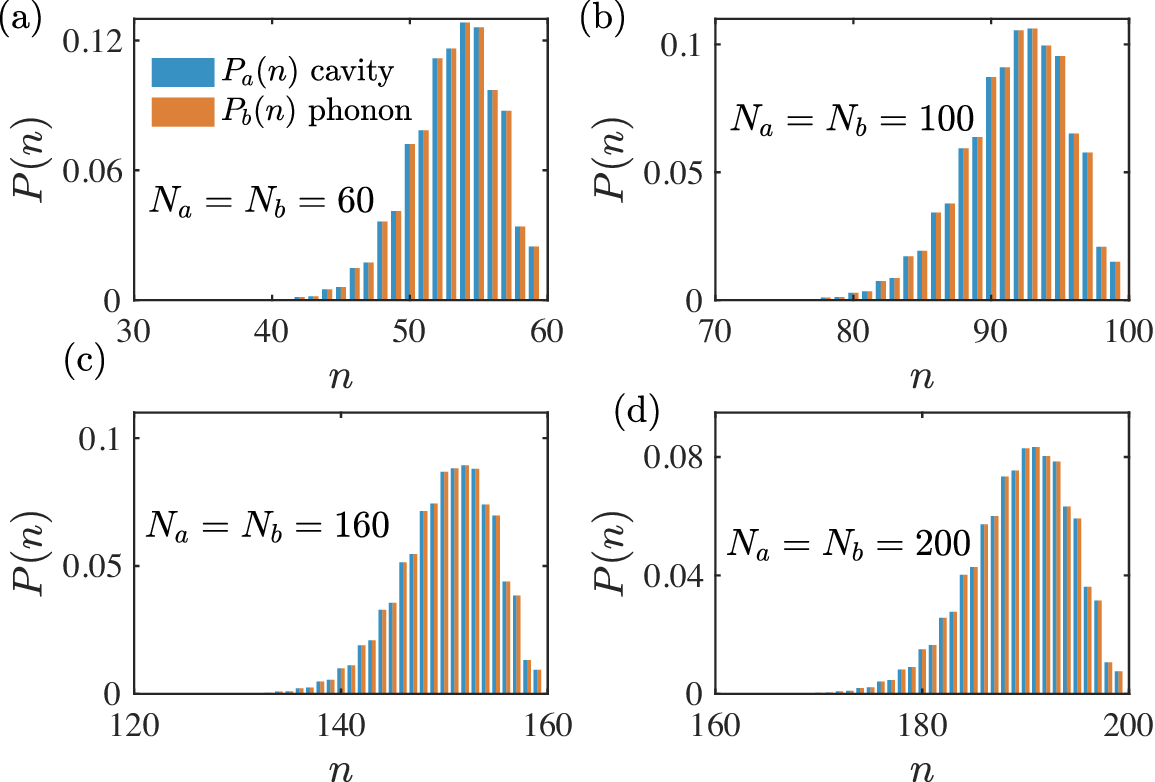}
\caption{Cutoff dependence of the ground-state Fock distributions in the Kerr-free limit ($K_0=0$). The cavity and phonon distributions are shown for increasing Hilbert-space truncations: (a) $N_a=N_b=60$, (b) $N_a=N_b=100$, (c) $N_a=N_b=160$, and (d) $N_a=N_b=200$. Parameters are $r_a=r_b=0.2$ and $\tilde g_1/\tilde\Delta_a=2$.
The probability maxima continuously shift toward the cutoff boundary as the truncation is increased, indicating the absence of a converged finite-occupation ground state in the Kerr-free model. This cutoff dependence reveals the unbounded growth of hybrid photon--phonon excitations when no nonlinear saturation mechanism is present.
}
\label{cutoff}
\end{figure}

For a finite-occupation hybrid state, minimizing Eq.~(\ref{Eg_app}) with respect to $N_a$ and $N_b$ gives
\begin{align}
\tilde\Delta_a+2U_aN_a
&=
\frac{\tilde g_1^2N_b}
{\sqrt{
\Delta_s^2+4\tilde g_1^2N_aN_b
}},
\label{stationary_Na_app}
\\
\tilde\Delta_b
&=
\frac{\tilde g_1^2N_a}
{\sqrt{
\Delta_s^2+4\tilde g_1^2N_aN_b
}}.
\label{stationary_Nb_app}
\end{align}
Eliminating the common spin-mediated energy scale immediately yields
\begin{equation}
\frac{N_b}{N_a}
=
\frac{
\tilde\Delta_a+2U_aN_a
}{
\tilde\Delta_b
}.
\label{occupation_ratio_app}
\end{equation}
Equation~(\ref{occupation_ratio_app}) exposes the microscopic action of Kerr filtering on the ordered-state composition. Without Kerr nonlinearity, the semiclassical photon and phonon weights are locked by their effective detunings,
\begin{equation}
\frac{N_b}{N_a}
=
\frac{\tilde\Delta_a}{\tilde\Delta_b}.
\label{occupation_ratio_zeroK}
\end{equation}
A finite $U_a$, however, produces an occupation-dependent cavity shift,
\begin{equation}
\frac{N_b}{N_a}
=
\frac{\tilde\Delta_a}{\tilde\Delta_b}
+
\frac{2U_a}{\tilde\Delta_b}N_a,
\label{occupation_ratio_Kerr}
\end{equation}
so that increasing the collective excitation progressively increases the energetic cost of the photonic component. The hybrid state is therefore continuously reweighted toward the phononic sector, while the underlying spin-mediated photon-phonon conversion process remains intact.

Equations~(\ref{critical_general_root}) and (\ref{occupation_ratio_Kerr}) summarize the complementary roles of double squeezing and Kerr nonlinearity. The squeezing-enhanced tripartite interaction controls the onset of hybrid ordering through the lowest correlated pair manifold, where the Kerr energy vanishes. Once higher excitation manifolds become populated, the Kerr nonlinearity acts as a mode-selective nonlinear filter that suppresses photonic accumulation and reconstructs the internal composition of the ordered state. This separation between the generation of collective hybrid excitations and their nonlinear redistribution provides the microscopic basis of the Kerr-filtered superradiant phases discussed in the main text.

\section{Role of Kerr nonlinearity in stabilizing the hybrid ordered state\label{sm_IV}}

To further clarify the role of the Kerr nonlinearity, we examine the ground-state properties in the Kerr-free limit $K_0=0$. In this case, the effective tripartite interaction is no longer balanced by an occupation-dependent nonlinear energy cost. For sufficiently strong coupling, the pair-generation term continuously lowers the energy by increasing the correlated photon-phonon occupation.

Figure~\ref{cutoff} shows the ground-state Fock distributions obtained with different Hilbert-space truncations. The cavity and phonon distributions remain nearly identical, reflecting the absence of a mechanism that differentiates the two bosonic components. More importantly, the occupation maxima systematically move toward larger photon and phonon numbers as the cutoff is increased. Therefore, the numerical ground state does not approach a cutoff-independent distribution, but instead exhibits an instability toward increasing hybrid excitation.

This behavior originates from the competition between the tripartite pair-generation energy and the nonlinear saturation energy. When $K_0=0$, the latter is absent, and the correlated photon-phonon excitations can grow without bound. A finite Kerr interaction introduces the nonlinear energy penalty $U_aN_a^2$, which stabilizes the ordered phase and enables the Kerr-induced reconstruction of the hybrid-state composition discussed in the main text.


\begin{thebibliography}{68}%
\makeatletter
\providecommand \@ifxundefined [1]{%
 \@ifx{#1\undefined}
}%
\providecommand \@ifnum [1]{%
 \ifnum #1\expandafter \@firstoftwo
 \else \expandafter \@secondoftwo
 \fi
}%
\providecommand \@ifx [1]{%
 \ifx #1\expandafter \@firstoftwo
 \else \expandafter \@secondoftwo
 \fi
}%
\providecommand \natexlab [1]{#1}%
\providecommand \enquote  [1]{``#1''}%
\providecommand \bibnamefont  [1]{#1}%
\providecommand \bibfnamefont [1]{#1}%
\providecommand \citenamefont [1]{#1}%
\providecommand \href@noop [0]{\@secondoftwo}%
\providecommand \href [0]{\begingroup \@sanitize@url \@href}%
\providecommand \@href[1]{\@@startlink{#1}\@@href}%
\providecommand \@@href[1]{\endgroup#1\@@endlink}%
\providecommand \@sanitize@url [0]{\catcode `\\12\catcode `\$12\catcode
  `\&12\catcode `\#12\catcode `\^12\catcode `\_12\catcode `\%12\relax}%
\providecommand \@@startlink[1]{}%
\providecommand \@@endlink[0]{}%
\providecommand \url  [0]{\begingroup\@sanitize@url \@url }%
\providecommand \@url [1]{\endgroup\@href {#1}{\urlprefix }}%
\providecommand \urlprefix  [0]{URL }%
\providecommand \Eprint [0]{\href }%
\providecommand \doibase [0]{http://dx.doi.org/}%
\providecommand \selectlanguage [0]{\@gobble}%
\providecommand \bibinfo  [0]{\@secondoftwo}%
\providecommand \bibfield  [0]{\@secondoftwo}%
\providecommand \translation [1]{[#1]}%
\providecommand \BibitemOpen [0]{}%
\providecommand \bibitemStop [0]{}%
\providecommand \bibitemNoStop [0]{.\EOS\space}%
\providecommand \EOS [0]{\spacefactor3000\relax}%
\providecommand \BibitemShut  [1]{\csname bibitem#1\endcsname}%
\let\auto@bib@innerbib\@empty
\bibitem [{\citenamefont {Haroche}(2013)}]{Haroche2013}%
  \BibitemOpen
  \bibfield  {author} {\bibinfo {author} {\bibfnamefont {Serge}\ \bibnamefont
  {Haroche}},\ }\bibfield  {title} {\enquote {\bibinfo {title} {Nobel lecture:
  Controlling photons in a box and exploring the quantum to classical
  boundary},}\ }\href {\doibase 10.1103/RevModPhys.85.1083} {\bibfield
  {journal} {\bibinfo  {journal} {Rev. Mod. Phys.}\ }\textbf {\bibinfo {volume}
  {85}},\ \bibinfo {pages} {1083--1102} (\bibinfo {year} {2013})}\BibitemShut
  {NoStop}%
\bibitem [{\citenamefont {Aspelmeyer}\ \emph {et~al.}(2014)\citenamefont
  {Aspelmeyer}, \citenamefont {Kippenberg},\ and\ \citenamefont
  {Marquardt}}]{Aspelmeyer2014}%
  \BibitemOpen
  \bibfield  {author} {\bibinfo {author} {\bibfnamefont {M.}~\bibnamefont
  {Aspelmeyer}}, \bibinfo {author} {\bibfnamefont {T.~J.}\ \bibnamefont
  {Kippenberg}}, \ and\ \bibinfo {author} {\bibfnamefont {F.}~\bibnamefont
  {Marquardt}},\ }\bibfield  {title} {\enquote {\bibinfo {title} {Cavity
  optomechanics},}\ }\href@noop {} {\bibfield  {journal} {\bibinfo  {journal}
  {Rev. Mod. Phys.}\ }\textbf {\bibinfo {volume} {86}},\ \bibinfo {pages}
  {1391} (\bibinfo {year} {2014})}\BibitemShut {NoStop}%
\bibitem [{\citenamefont {Blais}\ \emph {et~al.}(2021)\citenamefont {Blais},
  \citenamefont {Grimsmo}, \citenamefont {Girvin},\ and\ \citenamefont
  {Wallraff}}]{Blais2021}%
  \BibitemOpen
  \bibfield  {author} {\bibinfo {author} {\bibfnamefont {A.}~\bibnamefont
  {Blais}}, \bibinfo {author} {\bibfnamefont {A.~L.}\ \bibnamefont {Grimsmo}},
  \bibinfo {author} {\bibfnamefont {S.~M.}\ \bibnamefont {Girvin}}, \ and\
  \bibinfo {author} {\bibfnamefont {A.}~\bibnamefont {Wallraff}},\ }\bibfield
  {title} {\enquote {\bibinfo {title} {Circuit quantum electrodynamics},}\
  }\href@noop {} {\bibfield  {journal} {\bibinfo  {journal} {Rev. Mod. Phys.}\
  }\textbf {\bibinfo {volume} {93}},\ \bibinfo {pages} {025005} (\bibinfo
  {year} {2021})}\BibitemShut {NoStop}%
\bibitem [{\citenamefont {Arute}\ \emph {et~al.}(2019)\citenamefont {Arute},
  \citenamefont {Arya}, \citenamefont {Babbush}, \citenamefont {Bacon},
  \citenamefont {Bardin}, \citenamefont {Barends}, \citenamefont {Biswas},
  \citenamefont {Boixo}, \citenamefont {Brandao}, \citenamefont {Buell} \emph
  {et~al.}}]{Arute2019}%
  \BibitemOpen
  \bibfield  {author} {\bibinfo {author} {\bibfnamefont {Frank}\ \bibnamefont
  {Arute}}, \bibinfo {author} {\bibfnamefont {Kunal}\ \bibnamefont {Arya}},
  \bibinfo {author} {\bibfnamefont {Ryan}\ \bibnamefont {Babbush}}, \bibinfo
  {author} {\bibfnamefont {Dave}\ \bibnamefont {Bacon}}, \bibinfo {author}
  {\bibfnamefont {Joseph~C}\ \bibnamefont {Bardin}}, \bibinfo {author}
  {\bibfnamefont {Rami}\ \bibnamefont {Barends}}, \bibinfo {author}
  {\bibfnamefont {Rupak}\ \bibnamefont {Biswas}}, \bibinfo {author}
  {\bibfnamefont {Sergio}\ \bibnamefont {Boixo}}, \bibinfo {author}
  {\bibfnamefont {Fernando~GSL}\ \bibnamefont {Brandao}}, \bibinfo {author}
  {\bibfnamefont {David~A}\ \bibnamefont {Buell}},  \emph {et~al.},\ }\bibfield
   {title} {\enquote {\bibinfo {title} {Quantum supremacy using a programmable
  superconducting processor},}\ }\href@noop {} {\bibfield  {journal} {\bibinfo
  {journal} {nature}\ }\textbf {\bibinfo {volume} {574}},\ \bibinfo {pages}
  {505--510} (\bibinfo {year} {2019})}\BibitemShut {NoStop}%
\bibitem [{\citenamefont {Clerk}\ \emph {et~al.}(2010)\citenamefont {Clerk},
  \citenamefont {Devoret}, \citenamefont {Girvin}, \citenamefont {Marquardt},\
  and\ \citenamefont {Schoelkopf}}]{Clerk2010}%
  \BibitemOpen
  \bibfield  {author} {\bibinfo {author} {\bibfnamefont {A.~A.}\ \bibnamefont
  {Clerk}}, \bibinfo {author} {\bibfnamefont {M.~H.}\ \bibnamefont {Devoret}},
  \bibinfo {author} {\bibfnamefont {S.~M.}\ \bibnamefont {Girvin}}, \bibinfo
  {author} {\bibfnamefont {F.}~\bibnamefont {Marquardt}}, \ and\ \bibinfo
  {author} {\bibfnamefont {R.~J.}\ \bibnamefont {Schoelkopf}},\ }\bibfield
  {title} {\enquote {\bibinfo {title} {Introduction to quantum noise,
  measurement, and amplification},}\ }\href {\doibase
  10.1103/RevModPhys.82.1155} {\bibfield  {journal} {\bibinfo  {journal} {Rev.
  Mod. Phys.}\ }\textbf {\bibinfo {volume} {82}},\ \bibinfo {pages}
  {1155--1208} (\bibinfo {year} {2010})}\BibitemShut {NoStop}%
\bibitem [{\citenamefont {Kurizki}\ \emph {et~al.}(2015)\citenamefont
  {Kurizki}, \citenamefont {Bertet}, \citenamefont {Kubo}, \citenamefont
  {M{\o}lmer}, \citenamefont {Petrosyan}, \citenamefont {Rabl},\ and\
  \citenamefont {Schmiedmayer}}]{Kurizki2015}%
  \BibitemOpen
  \bibfield  {author} {\bibinfo {author} {\bibfnamefont {Gershon}\ \bibnamefont
  {Kurizki}}, \bibinfo {author} {\bibfnamefont {Patrice}\ \bibnamefont
  {Bertet}}, \bibinfo {author} {\bibfnamefont {Yuimaru}\ \bibnamefont {Kubo}},
  \bibinfo {author} {\bibfnamefont {Klaus}\ \bibnamefont {M{\o}lmer}}, \bibinfo
  {author} {\bibfnamefont {David}\ \bibnamefont {Petrosyan}}, \bibinfo {author}
  {\bibfnamefont {Peter}\ \bibnamefont {Rabl}}, \ and\ \bibinfo {author}
  {\bibfnamefont {J{\"o}rg}\ \bibnamefont {Schmiedmayer}},\ }\bibfield  {title}
  {\enquote {\bibinfo {title} {Quantum technologies with hybrid systems},}\
  }\href {\doibase 10.1073/pnas.1419326112} {\bibfield  {journal} {\bibinfo
  {journal} {Proc. Natl. Acad. Sci. U.S.A.}\ }\textbf {\bibinfo {volume}
  {112}},\ \bibinfo {pages} {3866--3873} (\bibinfo {year} {2015})}\BibitemShut
  {NoStop}%
\bibitem [{\citenamefont {Lauk}\ \emph {et~al.}(2020)\citenamefont {Lauk},
  \citenamefont {Sinclair}, \citenamefont {Barzanjeh}, \citenamefont {Covey},
  \citenamefont {Saffman}, \citenamefont {Spiropulu},\ and\ \citenamefont
  {Simon}}]{Lauk2020}%
  \BibitemOpen
  \bibfield  {author} {\bibinfo {author} {\bibfnamefont {Nikolai}\ \bibnamefont
  {Lauk}}, \bibinfo {author} {\bibfnamefont {Neil}\ \bibnamefont {Sinclair}},
  \bibinfo {author} {\bibfnamefont {Shabir}\ \bibnamefont {Barzanjeh}},
  \bibinfo {author} {\bibfnamefont {Jacob~P.}\ \bibnamefont {Covey}}, \bibinfo
  {author} {\bibfnamefont {Mark}\ \bibnamefont {Saffman}}, \bibinfo {author}
  {\bibfnamefont {Maria}\ \bibnamefont {Spiropulu}}, \ and\ \bibinfo {author}
  {\bibfnamefont {Christoph}\ \bibnamefont {Simon}},\ }\bibfield  {title}
  {\enquote {\bibinfo {title} {Perspectives on quantum transduction},}\ }\href
  {\doibase 10.1088/2058-9565/ab788a} {\bibfield  {journal} {\bibinfo
  {journal} {Quantum Sci. Technol.}\ }\textbf {\bibinfo {volume} {5}},\
  \bibinfo {pages} {020501} (\bibinfo {year} {2020})}\BibitemShut {NoStop}%
\bibitem [{\citenamefont {Rabi}(1936)}]{rabi1936process}%
  \BibitemOpen
  \bibfield  {author} {\bibinfo {author} {\bibfnamefont {II}~\bibnamefont
  {Rabi}},\ }\bibfield  {title} {\enquote {\bibinfo {title} {On the process of
  space quantization},}\ }\href@noop {} {\bibfield  {journal} {\bibinfo
  {journal} {Physical Review}\ }\textbf {\bibinfo {volume} {49}},\ \bibinfo
  {pages} {324} (\bibinfo {year} {1936})}\BibitemShut {NoStop}%
\bibitem [{\citenamefont {Jaynes}\ and\ \citenamefont
  {Cummings}(1963)}]{jaynes1963comparison}%
  \BibitemOpen
  \bibfield  {author} {\bibinfo {author} {\bibfnamefont {Edwin~T}\ \bibnamefont
  {Jaynes}}\ and\ \bibinfo {author} {\bibfnamefont {Frederick~W}\ \bibnamefont
  {Cummings}},\ }\bibfield  {title} {\enquote {\bibinfo {title} {Comparison of
  quantum and semiclassical radiation theories with application to the beam
  maser},}\ }\href@noop {} {\bibfield  {journal} {\bibinfo  {journal}
  {Proceedings of the IEEE}\ }\textbf {\bibinfo {volume} {51}},\ \bibinfo
  {pages} {89--109} (\bibinfo {year} {1963})}\BibitemShut {NoStop}%
\bibitem [{\citenamefont {Xiang}\ \emph {et~al.}(2013)\citenamefont {Xiang},
  \citenamefont {Ashhab}, \citenamefont {You},\ and\ \citenamefont
  {Nori}}]{xiang2013hybrid}%
  \BibitemOpen
  \bibfield  {author} {\bibinfo {author} {\bibfnamefont {Ze-Liang}\
  \bibnamefont {Xiang}}, \bibinfo {author} {\bibfnamefont {Sahel}\ \bibnamefont
  {Ashhab}}, \bibinfo {author} {\bibfnamefont {JQ}~\bibnamefont {You}}, \ and\
  \bibinfo {author} {\bibfnamefont {Franco}\ \bibnamefont {Nori}},\ }\bibfield
  {title} {\enquote {\bibinfo {title} {Hybrid quantum circuits: Superconducting
  circuits interacting with other quantum systems},}\ }\href@noop {} {\bibfield
   {journal} {\bibinfo  {journal} {Reviews of Modern Physics}\ }\textbf
  {\bibinfo {volume} {85}},\ \bibinfo {pages} {623--653} (\bibinfo {year}
  {2013})}\BibitemShut {NoStop}%
\bibitem [{\citenamefont {Shandilya}\ \emph {et~al.}(2021)\citenamefont
  {Shandilya}, \citenamefont {Lake}, \citenamefont {Mitchell}, \citenamefont
  {Sukachev},\ and\ \citenamefont {Barclay}}]{shandilya2021optomechanical}%
  \BibitemOpen
  \bibfield  {author} {\bibinfo {author} {\bibfnamefont {Prasoon~K}\
  \bibnamefont {Shandilya}}, \bibinfo {author} {\bibfnamefont {David~P}\
  \bibnamefont {Lake}}, \bibinfo {author} {\bibfnamefont {Matthew~J}\
  \bibnamefont {Mitchell}}, \bibinfo {author} {\bibfnamefont {Denis~D}\
  \bibnamefont {Sukachev}}, \ and\ \bibinfo {author} {\bibfnamefont {Paul~E}\
  \bibnamefont {Barclay}},\ }\bibfield  {title} {\enquote {\bibinfo {title}
  {Optomechanical interface between telecom photons and spin quantum memory},}\
  }\href@noop {} {\bibfield  {journal} {\bibinfo  {journal} {Nature Physics}\
  }\textbf {\bibinfo {volume} {17}},\ \bibinfo {pages} {1420--1425} (\bibinfo
  {year} {2021})}\BibitemShut {NoStop}%
\bibitem [{\citenamefont {Hei}\ \emph {et~al.}(2023)\citenamefont {Hei},
  \citenamefont {Li}, \citenamefont {Pan},\ and\ \citenamefont
  {Nori}}]{Hei2023}%
  \BibitemOpen
  \bibfield  {author} {\bibinfo {author} {\bibfnamefont {Xin-Lei}\ \bibnamefont
  {Hei}}, \bibinfo {author} {\bibfnamefont {Peng-Bo}\ \bibnamefont {Li}},
  \bibinfo {author} {\bibfnamefont {Xue-Feng}\ \bibnamefont {Pan}}, \ and\
  \bibinfo {author} {\bibfnamefont {Franco}\ \bibnamefont {Nori}},\ }\bibfield
  {title} {\enquote {\bibinfo {title} {Enhanced tripartite interactions in
  spin--magnon--mechanical hybrid systems},}\ }\href {\doibase
  10.1103/PhysRevLett.130.073602} {\bibfield  {journal} {\bibinfo  {journal}
  {Phys. Rev. Lett.}\ }\textbf {\bibinfo {volume} {130}},\ \bibinfo {pages}
  {073602} (\bibinfo {year} {2023})}\BibitemShut {NoStop}%
\bibitem [{\citenamefont {Tang}\ and\ \citenamefont
  {Deng}(2024)}]{Tang2024APL}%
  \BibitemOpen
  \bibfield  {author} {\bibinfo {author} {\bibfnamefont {Jing}\ \bibnamefont
  {Tang}}\ and\ \bibinfo {author} {\bibfnamefont {Yuangang}\ \bibnamefont
  {Deng}},\ }\bibfield  {title} {\enquote {\bibinfo {title} {Unveiling vacuum
  fluctuations and nonclassical states with cavity-enhanced tripartite
  interactions},}\ }\href {\doibase 10.1063/5.0210936} {\bibfield  {journal}
  {\bibinfo  {journal} {APL Photonics}\ }\textbf {\bibinfo {volume} {9}},\
  \bibinfo {pages} {096117} (\bibinfo {year} {2024})}\BibitemShut {NoStop}%
\bibitem [{\citenamefont {Lemonde}\ \emph
  {et~al.}(2016{\natexlab{a}})\citenamefont {Lemonde}, \citenamefont {Didier},\
  and\ \citenamefont {Clerk}}]{lemonde2016enhanced}%
  \BibitemOpen
  \bibfield  {author} {\bibinfo {author} {\bibfnamefont {Marc-Antoine}\
  \bibnamefont {Lemonde}}, \bibinfo {author} {\bibfnamefont {Nicolas}\
  \bibnamefont {Didier}}, \ and\ \bibinfo {author} {\bibfnamefont {Aashish~A}\
  \bibnamefont {Clerk}},\ }\bibfield  {title} {\enquote {\bibinfo {title}
  {Enhanced nonlinear interactions in quantum optomechanics via mechanical
  amplification},}\ }\href@noop {} {\bibfield  {journal} {\bibinfo  {journal}
  {Nature communications}\ }\textbf {\bibinfo {volume} {7}},\ \bibinfo {pages}
  {11338} (\bibinfo {year} {2016}{\natexlab{a}})}\BibitemShut {NoStop}%
\bibitem [{\citenamefont {Riedinger}\ and\ \citenamefont
  {et~al.}(2018)}]{Riedinger2018}%
  \BibitemOpen
  \bibfield  {author} {\bibinfo {author} {\bibfnamefont {R.}~\bibnamefont
  {Riedinger}}\ and\ \bibinfo {author} {\bibnamefont {et~al.}},\ }\bibfield
  {title} {\enquote {\bibinfo {title} {Remote quantum entanglement between two
  micromechanical oscillators},}\ }\href@noop {} {\bibfield  {journal}
  {\bibinfo  {journal} {Nature}\ }\textbf {\bibinfo {volume} {556}},\ \bibinfo
  {pages} {473} (\bibinfo {year} {2018})}\BibitemShut {NoStop}%
\bibitem [{\citenamefont {Higginbotham}\ \emph {et~al.}(2018)\citenamefont
  {Higginbotham} \emph {et~al.}}]{Higginbotham2018}%
  \BibitemOpen
  \bibfield  {author} {\bibinfo {author} {\bibfnamefont {A.~P.}\ \bibnamefont
  {Higginbotham}} \emph {et~al.},\ }\bibfield  {title} {\enquote {\bibinfo
  {title} {Harnessing electro-optic correlations in an efficient mechanical
  converter},}\ }\href {\doibase 10.1038/s41567-018-0205-0} {\bibfield
  {journal} {\bibinfo  {journal} {Nature Physics}\ }\textbf {\bibinfo {volume}
  {14}},\ \bibinfo {pages} {1038--1042} (\bibinfo {year} {2018})}\BibitemShut
  {NoStop}%
\bibitem [{\citenamefont {Hong}\ \emph {et~al.}(1987)\citenamefont {Hong},
  \citenamefont {Ou},\ and\ \citenamefont {Mandel}}]{Hong1987}%
  \BibitemOpen
  \bibfield  {author} {\bibinfo {author} {\bibfnamefont {C.~K.}\ \bibnamefont
  {Hong}}, \bibinfo {author} {\bibfnamefont {Z.~Y.}\ \bibnamefont {Ou}}, \ and\
  \bibinfo {author} {\bibfnamefont {L.}~\bibnamefont {Mandel}},\ }\bibfield
  {title} {\enquote {\bibinfo {title} {Measurement of subpicosecond time
  intervals between two photons by interference},}\ }\href {\doibase
  10.1103/PhysRevLett.59.2044} {\bibfield  {journal} {\bibinfo  {journal}
  {Phys. Rev. Lett.}\ }\textbf {\bibinfo {volume} {59}},\ \bibinfo {pages}
  {2044--2046} (\bibinfo {year} {1987})}\BibitemShut {NoStop}%
\bibitem [{\citenamefont {Grimm}\ \emph
  {et~al.}(2020{\natexlab{a}})\citenamefont {Grimm}, \citenamefont {Frattini},
  \citenamefont {Puri}, \citenamefont {Mundhada}, \citenamefont {Touzard},
  \citenamefont {Mirrahimi}, \citenamefont {Girvin}, \citenamefont {Shankar},\
  and\ \citenamefont {Devoret}}]{Grimm2020}%
  \BibitemOpen
  \bibfield  {author} {\bibinfo {author} {\bibfnamefont {Alexander}\
  \bibnamefont {Grimm}}, \bibinfo {author} {\bibfnamefont {N.~E.}\ \bibnamefont
  {Frattini}}, \bibinfo {author} {\bibfnamefont {Shruti}\ \bibnamefont {Puri}},
  \bibinfo {author} {\bibfnamefont {S.~O.}\ \bibnamefont {Mundhada}}, \bibinfo
  {author} {\bibfnamefont {Steven}\ \bibnamefont {Touzard}}, \bibinfo {author}
  {\bibfnamefont {Mazyar}\ \bibnamefont {Mirrahimi}}, \bibinfo {author}
  {\bibfnamefont {S.~M.}\ \bibnamefont {Girvin}}, \bibinfo {author}
  {\bibfnamefont {S.}~\bibnamefont {Shankar}}, \ and\ \bibinfo {author}
  {\bibfnamefont {M.~H.}\ \bibnamefont {Devoret}},\ }\bibfield  {title}
  {\enquote {\bibinfo {title} {Stabilization and operation of a {Kerr}-cat
  qubit},}\ }\href {\doibase 10.1038/s41586-020-2587-z} {\bibfield  {journal}
  {\bibinfo  {journal} {Nature}\ }\textbf {\bibinfo {volume} {584}},\ \bibinfo
  {pages} {205--209} (\bibinfo {year} {2020}{\natexlab{a}})}\BibitemShut
  {NoStop}%
\bibitem [{\citenamefont {Ockeloen-Korppi}\ and\ \citenamefont
  {et~al.}(2018)}]{OckeloenKorppi2018}%
  \BibitemOpen
  \bibfield  {author} {\bibinfo {author} {\bibfnamefont {C.~F.}\ \bibnamefont
  {Ockeloen-Korppi}}\ and\ \bibinfo {author} {\bibnamefont {et~al.}},\
  }\bibfield  {title} {\enquote {\bibinfo {title} {Stabilized entanglement of
  massive mechanical oscillators},}\ }\href@noop {} {\bibfield  {journal}
  {\bibinfo  {journal} {Nature}\ }\textbf {\bibinfo {volume} {556}},\ \bibinfo
  {pages} {478} (\bibinfo {year} {2018})}\BibitemShut {NoStop}%
\bibitem [{\citenamefont {Chu}\ \emph {et~al.}(2018)\citenamefont {Chu},
  \citenamefont {Kharel}, \citenamefont {Renninger}, \citenamefont {Burkhart},
  \citenamefont {Frunzio}, \citenamefont {Rakich},\ and\ \citenamefont
  {Schoelkopf}}]{Chu2018}%
  \BibitemOpen
  \bibfield  {author} {\bibinfo {author} {\bibfnamefont {Yiwen}\ \bibnamefont
  {Chu}}, \bibinfo {author} {\bibfnamefont {Praful}\ \bibnamefont {Kharel}},
  \bibinfo {author} {\bibfnamefont {William~H.}\ \bibnamefont {Renninger}},
  \bibinfo {author} {\bibfnamefont {Larry~D.}\ \bibnamefont {Burkhart}},
  \bibinfo {author} {\bibfnamefont {Luigi}\ \bibnamefont {Frunzio}}, \bibinfo
  {author} {\bibfnamefont {Peter~T.}\ \bibnamefont {Rakich}}, \ and\ \bibinfo
  {author} {\bibfnamefont {Robert~J.}\ \bibnamefont {Schoelkopf}},\ }\bibfield
  {title} {\enquote {\bibinfo {title} {Creation and control of multi-phonon
  fock states in a bulk acoustic-wave resonator},}\ }\href {\doibase
  10.1038/s41586-018-0691-z} {\bibfield  {journal} {\bibinfo  {journal}
  {Nature}\ }\textbf {\bibinfo {volume} {563}},\ \bibinfo {pages} {666--670}
  (\bibinfo {year} {2018})}\BibitemShut {NoStop}%
\bibitem [{\citenamefont {Forsch}\ and\ \citenamefont
  {et~al.}(2020)}]{Forsch2020}%
  \BibitemOpen
  \bibfield  {author} {\bibinfo {author} {\bibfnamefont {M.}~\bibnamefont
  {Forsch}}\ and\ \bibinfo {author} {\bibnamefont {et~al.}},\ }\bibfield
  {title} {\enquote {\bibinfo {title} {Microwave-to-optics conversion using a
  mechanical oscillator},}\ }\href@noop {} {\bibfield  {journal} {\bibinfo
  {journal} {Nature Physics}\ }\textbf {\bibinfo {volume} {16}},\ \bibinfo
  {pages} {69} (\bibinfo {year} {2020})}\BibitemShut {NoStop}%
\bibitem [{\citenamefont {Andrews}\ \emph {et~al.}(2014)\citenamefont
  {Andrews}, \citenamefont {Peterson}, \citenamefont {Purdy}, \citenamefont
  {Cicak}, \citenamefont {Simmonds}, \citenamefont {Regal},\ and\ \citenamefont
  {Lehnert}}]{Andrews2014}%
  \BibitemOpen
  \bibfield  {author} {\bibinfo {author} {\bibfnamefont {R.~W.}\ \bibnamefont
  {Andrews}}, \bibinfo {author} {\bibfnamefont {R.~W.}\ \bibnamefont
  {Peterson}}, \bibinfo {author} {\bibfnamefont {T.~P.}\ \bibnamefont {Purdy}},
  \bibinfo {author} {\bibfnamefont {K.}~\bibnamefont {Cicak}}, \bibinfo
  {author} {\bibfnamefont {R.~W.}\ \bibnamefont {Simmonds}}, \bibinfo {author}
  {\bibfnamefont {C.~A.}\ \bibnamefont {Regal}}, \ and\ \bibinfo {author}
  {\bibfnamefont {K.~W.}\ \bibnamefont {Lehnert}},\ }\bibfield  {title}
  {\enquote {\bibinfo {title} {Bidirectional and efficient conversion between
  microwave and optical light},}\ }\href {\doibase 10.1038/nphys2911}
  {\bibfield  {journal} {\bibinfo  {journal} {Nature Physics}\ }\textbf
  {\bibinfo {volume} {10}},\ \bibinfo {pages} {321--326} (\bibinfo {year}
  {2014})}\BibitemShut {NoStop}%
\bibitem [{\citenamefont {Riedinger}\ \emph {et~al.}(2016)\citenamefont
  {Riedinger}, \citenamefont {Hong}, \citenamefont {Norte}, \citenamefont
  {Slater}, \citenamefont {Shang}, \citenamefont {Krause}, \citenamefont
  {Anant}, \citenamefont {Aspelmeyer},\ and\ \citenamefont
  {Gr{\"o}blacher}}]{Riedinger2016}%
  \BibitemOpen
  \bibfield  {author} {\bibinfo {author} {\bibfnamefont {Roman}\ \bibnamefont
  {Riedinger}}, \bibinfo {author} {\bibfnamefont {Seok}\ \bibnamefont {Hong}},
  \bibinfo {author} {\bibfnamefont {Richard~A.}\ \bibnamefont {Norte}},
  \bibinfo {author} {\bibfnamefont {John~A.}\ \bibnamefont {Slater}}, \bibinfo
  {author} {\bibfnamefont {Jingxuan}\ \bibnamefont {Shang}}, \bibinfo {author}
  {\bibfnamefont {Alex~G.}\ \bibnamefont {Krause}}, \bibinfo {author}
  {\bibfnamefont {Vijay}\ \bibnamefont {Anant}}, \bibinfo {author}
  {\bibfnamefont {Markus}\ \bibnamefont {Aspelmeyer}}, \ and\ \bibinfo {author}
  {\bibfnamefont {Simon}\ \bibnamefont {Gr{\"o}blacher}},\ }\bibfield  {title}
  {\enquote {\bibinfo {title} {Non-classical correlations between single
  photons and phonons from a mechanical oscillator},}\ }\href {\doibase
  10.1038/nature16536} {\bibfield  {journal} {\bibinfo  {journal} {Nature}\
  }\textbf {\bibinfo {volume} {530}},\ \bibinfo {pages} {313--316} (\bibinfo
  {year} {2016})}\BibitemShut {NoStop}%
\bibitem [{\citenamefont {Safavi-Naeini}\ \emph {et~al.}(2013)\citenamefont
  {Safavi-Naeini} \emph {et~al.}}]{SafaviNaeini2013}%
  \BibitemOpen
  \bibfield  {author} {\bibinfo {author} {\bibfnamefont {Amir~H.}\ \bibnamefont
  {Safavi-Naeini}} \emph {et~al.},\ }\bibfield  {title} {\enquote {\bibinfo
  {title} {Squeezed light from a silicon micromechanical resonator},}\ }\href
  {\doibase 10.1038/nature12307} {\bibfield  {journal} {\bibinfo  {journal}
  {Nature}\ }\textbf {\bibinfo {volume} {500}},\ \bibinfo {pages} {185--189}
  (\bibinfo {year} {2013})}\BibitemShut {NoStop}%
\bibitem [{\citenamefont {Underwood}\ \emph {et~al.}(2015)\citenamefont
  {Underwood}, \citenamefont {Mason}, \citenamefont {Lee}, \citenamefont
  {Xuereb}, \citenamefont {Clerk},\ and\ \citenamefont
  {Harris}}]{Underwood2015}%
  \BibitemOpen
  \bibfield  {author} {\bibinfo {author} {\bibfnamefont {Maxwell}\ \bibnamefont
  {Underwood}}, \bibinfo {author} {\bibfnamefont {David}\ \bibnamefont
  {Mason}}, \bibinfo {author} {\bibfnamefont {Dongsung}\ \bibnamefont {Lee}},
  \bibinfo {author} {\bibfnamefont {Andre}\ \bibnamefont {Xuereb}}, \bibinfo
  {author} {\bibfnamefont {A.~A.}\ \bibnamefont {Clerk}}, \ and\ \bibinfo
  {author} {\bibfnamefont {J.~G.~E.}\ \bibnamefont {Harris}},\ }\bibfield
  {title} {\enquote {\bibinfo {title} {Measurement of the motional sidebands of
  a nanogram-scale oscillator in the quantum regime},}\ }\href {\doibase
  10.1103/PhysRevA.92.061801} {\bibfield  {journal} {\bibinfo  {journal} {Phys.
  Rev. A}\ }\textbf {\bibinfo {volume} {92}},\ \bibinfo {pages} {061801}
  (\bibinfo {year} {2015})}\BibitemShut {NoStop}%
\bibitem [{\citenamefont {Hong}\ \emph {et~al.}(2017)\citenamefont {Hong} \emph
  {et~al.}}]{Hong2017}%
  \BibitemOpen
  \bibfield  {author} {\bibinfo {author} {\bibfnamefont {Seok}\ \bibnamefont
  {Hong}} \emph {et~al.},\ }\bibfield  {title} {\enquote {\bibinfo {title}
  {Hanbury brown and twiss interferometry of single phonons from an
  optomechanical resonator},}\ }\href {\doibase 10.1126/science.aan7939}
  {\bibfield  {journal} {\bibinfo  {journal} {Science}\ }\textbf {\bibinfo
  {volume} {358}},\ \bibinfo {pages} {203--206} (\bibinfo {year}
  {2017})}\BibitemShut {NoStop}%
\bibitem [{\citenamefont {Liao}\ and\ \citenamefont {Nori}(2019)}]{Liao2019}%
  \BibitemOpen
  \bibfield  {author} {\bibinfo {author} {\bibfnamefont {Jie-Qiao}\
  \bibnamefont {Liao}}\ and\ \bibinfo {author} {\bibfnamefont {Franco}\
  \bibnamefont {Nori}},\ }\bibfield  {title} {\enquote {\bibinfo {title}
  {Photon blockade in quadratically coupled optomechanical systems enhanced by
  mechanical parametric amplification},}\ }\href {\doibase
  10.1103/PhysRevA.99.023838} {\bibfield  {journal} {\bibinfo  {journal} {Phys.
  Rev. A}\ }\textbf {\bibinfo {volume} {99}},\ \bibinfo {pages} {023838}
  (\bibinfo {year} {2019})}\BibitemShut {NoStop}%
\bibitem [{\citenamefont {Qiao}\ \emph {et~al.}(2026)\citenamefont {Qiao},
  \citenamefont {An},\ and\ \citenamefont {Li}}]{Qiao2026}%
  \BibitemOpen
  \bibfield  {author} {\bibinfo {author} {\bibfnamefont {Yi-Fan}\ \bibnamefont
  {Qiao}}, \bibinfo {author} {\bibfnamefont {Jun-Hong}\ \bibnamefont {An}}, \
  and\ \bibinfo {author} {\bibfnamefont {Peng-Bo}\ \bibnamefont {Li}},\
  }\bibfield  {title} {\enquote {\bibinfo {title} {Mechanical squeezed-fock
  qubit: Towards quantum weak-force sensing},}\ }\href@noop {} {\bibfield
  {journal} {\bibinfo  {journal} {Physical Review Letters}\ }\textbf {\bibinfo
  {volume} {136}},\ \bibinfo {pages} {040801} (\bibinfo {year}
  {2026})}\BibitemShut {NoStop}%
\bibitem [{\citenamefont {Lemonde}\ \emph
  {et~al.}(2016{\natexlab{b}})\citenamefont {Lemonde}, \citenamefont {Didier},\
  and\ \citenamefont {Clerk}}]{Lemonde2016}%
  \BibitemOpen
  \bibfield  {author} {\bibinfo {author} {\bibfnamefont {Marc-Antoine}\
  \bibnamefont {Lemonde}}, \bibinfo {author} {\bibfnamefont {Nicolas}\
  \bibnamefont {Didier}}, \ and\ \bibinfo {author} {\bibfnamefont {Aashish~A.}\
  \bibnamefont {Clerk}},\ }\bibfield  {title} {\enquote {\bibinfo {title}
  {Enhanced nonlinear interactions in quantum optomechanics via mechanical
  amplification},}\ }\href {\doibase 10.1038/ncomms11338} {\bibfield  {journal}
  {\bibinfo  {journal} {Nat. Commun.}\ }\textbf {\bibinfo {volume} {7}},\
  \bibinfo {pages} {11338} (\bibinfo {year} {2016}{\natexlab{b}})}\BibitemShut
  {NoStop}%
\bibitem [{\citenamefont {Qin}\ \emph {et~al.}(2018)\citenamefont {Qin},
  \citenamefont {Miranowicz}, \citenamefont {Li}, \citenamefont {L{\"u}},
  \citenamefont {You},\ and\ \citenamefont {Nori}}]{Qin2018}%
  \BibitemOpen
  \bibfield  {author} {\bibinfo {author} {\bibfnamefont {Wei}\ \bibnamefont
  {Qin}}, \bibinfo {author} {\bibfnamefont {Adam}\ \bibnamefont {Miranowicz}},
  \bibinfo {author} {\bibfnamefont {Peng-Bo}\ \bibnamefont {Li}}, \bibinfo
  {author} {\bibfnamefont {Xin-You}\ \bibnamefont {L{\"u}}}, \bibinfo {author}
  {\bibfnamefont {J.~Q.}\ \bibnamefont {You}}, \ and\ \bibinfo {author}
  {\bibfnamefont {Franco}\ \bibnamefont {Nori}},\ }\bibfield  {title} {\enquote
  {\bibinfo {title} {Exponentially enhanced light--matter interaction,
  cooperativities, and steady-state entanglement using parametric
  amplification},}\ }\href {\doibase 10.1103/PhysRevLett.120.093601} {\bibfield
   {journal} {\bibinfo  {journal} {Phys. Rev. Lett.}\ }\textbf {\bibinfo
  {volume} {120}},\ \bibinfo {pages} {093601} (\bibinfo {year}
  {2018})}\BibitemShut {NoStop}%
\bibitem [{\citenamefont {Wang}\ \emph {et~al.}(2022)\citenamefont {Wang},
  \citenamefont {Miranowicz},\ and\ \citenamefont {Nori}}]{Wang2022}%
  \BibitemOpen
  \bibfield  {author} {\bibinfo {author} {\bibfnamefont {Xin}\ \bibnamefont
  {Wang}}, \bibinfo {author} {\bibfnamefont {Adam}\ \bibnamefont {Miranowicz}},
  \ and\ \bibinfo {author} {\bibfnamefont {Franco}\ \bibnamefont {Nori}},\
  }\bibfield  {title} {\enquote {\bibinfo {title} {Generating strong quantum
  nonlinearities by parametric amplification},}\ }\href {\doibase
  10.1103/PhysRevLett.129.203601} {\bibfield  {journal} {\bibinfo  {journal}
  {Phys. Rev. Lett.}\ }\textbf {\bibinfo {volume} {129}},\ \bibinfo {pages}
  {203601} (\bibinfo {year} {2022})}\BibitemShut {NoStop}%
\bibitem [{\citenamefont {Didier}\ \emph {et~al.}(2014)\citenamefont {Didier},
  \citenamefont {Qassemi},\ and\ \citenamefont {Blais}}]{Didier2015}%
  \BibitemOpen
  \bibfield  {author} {\bibinfo {author} {\bibfnamefont {Nicolas}\ \bibnamefont
  {Didier}}, \bibinfo {author} {\bibfnamefont {Farzad}\ \bibnamefont
  {Qassemi}}, \ and\ \bibinfo {author} {\bibfnamefont {Alexandre}\ \bibnamefont
  {Blais}},\ }\bibfield  {title} {\enquote {\bibinfo {title} {Perfect squeezing
  by damping modulation in circuit quantum electrodynamics},}\ }\href@noop {}
  {\bibfield  {journal} {\bibinfo  {journal} {Phys. Rev. A}\ }\textbf {\bibinfo
  {volume} {89}},\ \bibinfo {pages} {013820} (\bibinfo {year}
  {2014})}\BibitemShut {NoStop}%
\bibitem [{\citenamefont {L{\"u}}\ \emph {et~al.}(2015)\citenamefont {L{\"u}},
  \citenamefont {Jing}, \citenamefont {Ma},\ and\ \citenamefont {Wu}}]{Lu2015}%
  \BibitemOpen
  \bibfield  {author} {\bibinfo {author} {\bibfnamefont {Xin-You}\ \bibnamefont
  {L{\"u}}}, \bibinfo {author} {\bibfnamefont {Hui}\ \bibnamefont {Jing}},
  \bibinfo {author} {\bibfnamefont {Jin-Yong}\ \bibnamefont {Ma}}, \ and\
  \bibinfo {author} {\bibfnamefont {Ying}\ \bibnamefont {Wu}},\ }\bibfield
  {title} {\enquote {\bibinfo {title} {Pt-symmetry-breaking chaos in
  optomechanics},}\ }\href@noop {} {\bibfield  {journal} {\bibinfo  {journal}
  {Physical review letters}\ }\textbf {\bibinfo {volume} {114}},\ \bibinfo
  {pages} {253601} (\bibinfo {year} {2015})}\BibitemShut {NoStop}%
\bibitem [{\citenamefont {Zhang}\ and\ \citenamefont
  {et~al.}(2016)}]{Zhang2016}%
  \BibitemOpen
  \bibfield  {author} {\bibinfo {author} {\bibfnamefont {X.}~\bibnamefont
  {Zhang}}\ and\ \bibinfo {author} {\bibnamefont {et~al.}},\ }\bibfield
  {title} {\enquote {\bibinfo {title} {Cavity magnomechanics},}\ }\href@noop {}
  {\bibfield  {journal} {\bibinfo  {journal} {Science Advances}\ }\textbf
  {\bibinfo {volume} {2}},\ \bibinfo {pages} {e1501286} (\bibinfo {year}
  {2016})}\BibitemShut {NoStop}%
\bibitem [{\citenamefont {Li}\ and\ \citenamefont
  {et~al.}(2018)}]{Li2018Magnon}%
  \BibitemOpen
  \bibfield  {author} {\bibinfo {author} {\bibfnamefont {J.}~\bibnamefont
  {Li}}\ and\ \bibinfo {author} {\bibnamefont {et~al.}},\ }\bibfield  {title}
  {\enquote {\bibinfo {title} {Magnon kerr effect in cavity magnomechanics},}\
  }\href@noop {} {\bibfield  {journal} {\bibinfo  {journal} {Phys. Rev. Lett.}\
  }\textbf {\bibinfo {volume} {121}},\ \bibinfo {pages} {203601} (\bibinfo
  {year} {2018})}\BibitemShut {NoStop}%
\bibitem [{\citenamefont {Harder}\ \emph {et~al.}(2018)\citenamefont {Harder}
  \emph {et~al.}}]{Harder2018}%
  \BibitemOpen
  \bibfield  {author} {\bibinfo {author} {\bibfnamefont {M.}~\bibnamefont
  {Harder}} \emph {et~al.},\ }\bibfield  {title} {\enquote {\bibinfo {title}
  {Level attraction due to dissipative magnon-photon coupling},}\ }\href@noop
  {} {\bibfield  {journal} {\bibinfo  {journal} {Phys. Rev. Lett.}\ }\textbf
  {\bibinfo {volume} {121}},\ \bibinfo {pages} {137203} (\bibinfo {year}
  {2018})}\BibitemShut {NoStop}%
\bibitem [{\citenamefont {Frattini}\ and\ \citenamefont
  {et~al.}(2023)}]{Frattini2023}%
  \BibitemOpen
  \bibfield  {author} {\bibinfo {author} {\bibfnamefont {N.~E.}\ \bibnamefont
  {Frattini}}\ and\ \bibinfo {author} {\bibnamefont {et~al.}},\ }\bibfield
  {title} {\enquote {\bibinfo {title} {3-wave mixing josephson dipole
  element},}\ }\href@noop {} {\bibfield  {journal} {\bibinfo  {journal}
  {Nature}\ }\textbf {\bibinfo {volume} {618}},\ \bibinfo {pages} {264}
  (\bibinfo {year} {2023})}\BibitemShut {NoStop}%
\bibitem [{\citenamefont {Cai}\ \emph {et~al.}(2025)\citenamefont {Cai},
  \citenamefont {Deng}, \citenamefont {Zhang}, \citenamefont {Ni},
  \citenamefont {Mai}, \citenamefont {Huang}, \citenamefont {Zheng},
  \citenamefont {Hu}, \citenamefont {Liu}, \citenamefont {Xu} \emph
  {et~al.}}]{Cai2026}%
  \BibitemOpen
  \bibfield  {author} {\bibinfo {author} {\bibfnamefont {Yanyan}\ \bibnamefont
  {Cai}}, \bibinfo {author} {\bibfnamefont {Xiaowei}\ \bibnamefont {Deng}},
  \bibinfo {author} {\bibfnamefont {Libo}\ \bibnamefont {Zhang}}, \bibinfo
  {author} {\bibfnamefont {Zhongchu}\ \bibnamefont {Ni}}, \bibinfo {author}
  {\bibfnamefont {Jiasheng}\ \bibnamefont {Mai}}, \bibinfo {author}
  {\bibfnamefont {Peihao}\ \bibnamefont {Huang}}, \bibinfo {author}
  {\bibfnamefont {Pan}\ \bibnamefont {Zheng}}, \bibinfo {author} {\bibfnamefont
  {Ling}\ \bibnamefont {Hu}}, \bibinfo {author} {\bibfnamefont {Song}\
  \bibnamefont {Liu}}, \bibinfo {author} {\bibfnamefont {Yuan}\ \bibnamefont
  {Xu}},  \emph {et~al.},\ }\bibfield  {title} {\enquote {\bibinfo {title}
  {Quantum squeezing amplification with a weak kerr nonlinear oscillator},}\
  }\href@noop {} {\bibfield  {journal} {\bibinfo  {journal} {Nature
  Communications}\ }\textbf {\bibinfo {volume} {17}},\ \bibinfo {pages} {970}
  (\bibinfo {year} {2025})}\BibitemShut {NoStop}%
\bibitem [{\citenamefont {Meher}\ \emph {et~al.}(2026)\citenamefont {Meher},
  \citenamefont {Pathak},\ and\ \citenamefont {Sivakumar}}]{Meher2026}%
  \BibitemOpen
  \bibfield  {author} {\bibinfo {author} {\bibfnamefont {Nilakantha}\
  \bibnamefont {Meher}}, \bibinfo {author} {\bibfnamefont {Anirban}\
  \bibnamefont {Pathak}}, \ and\ \bibinfo {author} {\bibfnamefont
  {S.}~\bibnamefont {Sivakumar}},\ }\bibfield  {title} {\enquote {\bibinfo
  {title} {Generation of large fock states from coherent states using kerr
  interaction and displacement},}\ }\href@noop {} {\bibfield  {journal}
  {\bibinfo  {journal} {Phys. Rev. A}\ }\textbf {\bibinfo {volume} {113}},\
  \bibinfo {pages} {033704} (\bibinfo {year} {2026})}\BibitemShut {NoStop}%
\bibitem [{\citenamefont {Kronwald}\ \emph {et~al.}(2013)\citenamefont
  {Kronwald}, \citenamefont {Marquardt},\ and\ \citenamefont
  {Clerk}}]{Kronwald2013}%
  \BibitemOpen
  \bibfield  {author} {\bibinfo {author} {\bibfnamefont {Andreas}\ \bibnamefont
  {Kronwald}}, \bibinfo {author} {\bibfnamefont {Florian}\ \bibnamefont
  {Marquardt}}, \ and\ \bibinfo {author} {\bibfnamefont {A.~A.}\ \bibnamefont
  {Clerk}},\ }\bibfield  {title} {\enquote {\bibinfo {title} {Arbitrarily large
  steady-state bosonic squeezing via dissipation},}\ }\href {\doibase
  10.1103/PhysRevA.88.063833} {\bibfield  {journal} {\bibinfo  {journal} {Phys.
  Rev. A}\ }\textbf {\bibinfo {volume} {88}},\ \bibinfo {pages} {063833}
  (\bibinfo {year} {2013})}\BibitemShut {NoStop}%
\bibitem [{\citenamefont {Wollman}\ \emph {et~al.}(2015)\citenamefont {Wollman}
  \emph {et~al.}}]{Wollman2015}%
  \BibitemOpen
  \bibfield  {author} {\bibinfo {author} {\bibfnamefont {Emma~E.}\ \bibnamefont
  {Wollman}} \emph {et~al.},\ }\bibfield  {title} {\enquote {\bibinfo {title}
  {Quantum squeezing of motion in a mechanical resonator},}\ }\href {\doibase
  10.1126/science.aac5138} {\bibfield  {journal} {\bibinfo  {journal}
  {Science}\ }\textbf {\bibinfo {volume} {349}},\ \bibinfo {pages} {952--955}
  (\bibinfo {year} {2015})}\BibitemShut {NoStop}%
\bibitem [{\citenamefont {Lecocq}\ \emph {et~al.}(2015)\citenamefont {Lecocq},
  \citenamefont {Clark}, \citenamefont {Simmonds}, \citenamefont {Aumentado},\
  and\ \citenamefont {Teufel}}]{Lecocq2015}%
  \BibitemOpen
  \bibfield  {author} {\bibinfo {author} {\bibfnamefont {Florent}\ \bibnamefont
  {Lecocq}}, \bibinfo {author} {\bibfnamefont {J.~B.}\ \bibnamefont {Clark}},
  \bibinfo {author} {\bibfnamefont {R.~W.}\ \bibnamefont {Simmonds}}, \bibinfo
  {author} {\bibfnamefont {J.}~\bibnamefont {Aumentado}}, \ and\ \bibinfo
  {author} {\bibfnamefont {J.~D.}\ \bibnamefont {Teufel}},\ }\bibfield  {title}
  {\enquote {\bibinfo {title} {Quantum nondemolition measurement of a
  nonclassical state of a massive object},}\ }\href {\doibase
  10.1103/PhysRevX.5.041037} {\bibfield  {journal} {\bibinfo  {journal} {Phys.
  Rev. X}\ }\textbf {\bibinfo {volume} {5}},\ \bibinfo {pages} {041037}
  (\bibinfo {year} {2015})}\BibitemShut {NoStop}%
\bibitem [{\citenamefont {Levitan}\ \emph {et~al.}(2016)\citenamefont
  {Levitan}, \citenamefont {Metelmann},\ and\ \citenamefont
  {Clerk}}]{Levitan2016}%
  \BibitemOpen
  \bibfield  {author} {\bibinfo {author} {\bibfnamefont {B.~A.}\ \bibnamefont
  {Levitan}}, \bibinfo {author} {\bibfnamefont {A.}~\bibnamefont {Metelmann}},
  \ and\ \bibinfo {author} {\bibfnamefont {A.~A.}\ \bibnamefont {Clerk}},\
  }\bibfield  {title} {\enquote {\bibinfo {title} {Optomechanics with
  two-phonon driving},}\ }\href {\doibase 10.1103/PhysRevA.93.013803}
  {\bibfield  {journal} {\bibinfo  {journal} {Phys. Rev. A}\ }\textbf {\bibinfo
  {volume} {93}},\ \bibinfo {pages} {013803} (\bibinfo {year}
  {2016})}\BibitemShut {NoStop}%
\bibitem [{\citenamefont {Metelmann}\ and\ \citenamefont
  {Clerk}(2015)}]{Metelmann2017}%
  \BibitemOpen
  \bibfield  {author} {\bibinfo {author} {\bibfnamefont {A.}~\bibnamefont
  {Metelmann}}\ and\ \bibinfo {author} {\bibfnamefont {A.~A.}\ \bibnamefont
  {Clerk}},\ }\bibfield  {title} {\enquote {\bibinfo {title} {Nonreciprocal
  photon transmission and amplification via reservoir engineering},}\ }\href
  {\doibase 10.1103/PhysRevX.5.021025} {\bibfield  {journal} {\bibinfo
  {journal} {Phys. Rev. X}\ }\textbf {\bibinfo {volume} {5}},\ \bibinfo {pages}
  {021025} (\bibinfo {year} {2015})}\BibitemShut {NoStop}%
\bibitem [{\citenamefont {Malz}\ and\ \citenamefont
  {Nunnenkamp}(2020)}]{Malz2020}%
  \BibitemOpen
  \bibfield  {author} {\bibinfo {author} {\bibfnamefont {Daniel}\ \bibnamefont
  {Malz}}\ and\ \bibinfo {author} {\bibfnamefont {Andreas}\ \bibnamefont
  {Nunnenkamp}},\ }\bibfield  {title} {\enquote {\bibinfo {title} {Ground-state
  cooling and high-fidelity quantum transduction via parametrically driven
  bad-cavity optomechanics},}\ }\href {\doibase 10.1103/PhysRevLett.124.103602}
  {\bibfield  {journal} {\bibinfo  {journal} {Phys. Rev. Lett.}\ }\textbf
  {\bibinfo {volume} {124}},\ \bibinfo {pages} {103602} (\bibinfo {year}
  {2020})}\BibitemShut {NoStop}%
\bibitem [{\citenamefont {Liu}\ \emph {et~al.}(2013)\citenamefont {Liu},
  \citenamefont {Xiao}, \citenamefont {Chen}, \citenamefont {Yu},\ and\
  \citenamefont {Gong}}]{Liu2013PRL}%
  \BibitemOpen
  \bibfield  {author} {\bibinfo {author} {\bibfnamefont {Yong-Chun}\
  \bibnamefont {Liu}}, \bibinfo {author} {\bibfnamefont {Yun-Feng}\
  \bibnamefont {Xiao}}, \bibinfo {author} {\bibfnamefont {You-Ling}\
  \bibnamefont {Chen}}, \bibinfo {author} {\bibfnamefont {Xiao-Chong}\
  \bibnamefont {Yu}}, \ and\ \bibinfo {author} {\bibfnamefont {Qihuang}\
  \bibnamefont {Gong}},\ }\bibfield  {title} {\enquote {\bibinfo {title}
  {Parametric down-conversion and polariton pair generation in optomechanical
  systems},}\ }\href {\doibase 10.1103/PhysRevLett.111.083601} {\bibfield
  {journal} {\bibinfo  {journal} {Phys. Rev. Lett.}\ }\textbf {\bibinfo
  {volume} {111}},\ \bibinfo {pages} {083601} (\bibinfo {year}
  {2013})}\BibitemShut {NoStop}%
\bibitem [{\citenamefont {Galland}\ \emph {et~al.}(2014)\citenamefont
  {Galland}, \citenamefont {Sangouard}, \citenamefont {Piro}, \citenamefont
  {Gisin},\ and\ \citenamefont {Kippenberg}}]{Galland2014PRL}%
  \BibitemOpen
  \bibfield  {author} {\bibinfo {author} {\bibfnamefont {Christophe}\
  \bibnamefont {Galland}}, \bibinfo {author} {\bibfnamefont {Nicolas}\
  \bibnamefont {Sangouard}}, \bibinfo {author} {\bibfnamefont {Nicolas}\
  \bibnamefont {Piro}}, \bibinfo {author} {\bibfnamefont {Nicolas}\
  \bibnamefont {Gisin}}, \ and\ \bibinfo {author} {\bibfnamefont {Tobias~J.}\
  \bibnamefont {Kippenberg}},\ }\bibfield  {title} {\enquote {\bibinfo {title}
  {Heralded single-phonon preparation, storage, and readout in cavity
  optomechanics},}\ }\href {\doibase 10.1103/PhysRevLett.112.143602} {\bibfield
   {journal} {\bibinfo  {journal} {Phys. Rev. Lett.}\ }\textbf {\bibinfo
  {volume} {112}},\ \bibinfo {pages} {143602} (\bibinfo {year}
  {2014})}\BibitemShut {NoStop}%
\bibitem [{\citenamefont {Greve}\ \emph {et~al.}(2021)\citenamefont {Greve}
  \emph {et~al.}}]{Greve2021Nature}%
  \BibitemOpen
  \bibfield  {author} {\bibinfo {author} {\bibfnamefont {Kevin~De}\
  \bibnamefont {Greve}} \emph {et~al.},\ }\bibfield  {title} {\enquote
  {\bibinfo {title} {Quantum-dot spin--phonon--photon interface for quantum
  networks},}\ }\href {\doibase 10.1038/s41586-021-04179-6} {\bibfield
  {journal} {\bibinfo  {journal} {Nature}\ }\textbf {\bibinfo {volume} {600}},\
  \bibinfo {pages} {667--671} (\bibinfo {year} {2021})}\BibitemShut {NoStop}%
\bibitem [{\citenamefont {Mayor}\ \emph {et~al.}(2025)\citenamefont {Mayor},
  \citenamefont {Malik}, \citenamefont {Primo}, \citenamefont {Gyger},
  \citenamefont {Jiang}, \citenamefont {Alegre},\ and\ \citenamefont
  {Safavi-Naeini}}]{Mayor2025NatCommun}%
  \BibitemOpen
  \bibfield  {author} {\bibinfo {author} {\bibfnamefont {Felix~M.}\
  \bibnamefont {Mayor}}, \bibinfo {author} {\bibfnamefont {Sultan}\
  \bibnamefont {Malik}}, \bibinfo {author} {\bibfnamefont {Andr{\'e}~G.}\
  \bibnamefont {Primo}}, \bibinfo {author} {\bibfnamefont {Samuel}\
  \bibnamefont {Gyger}}, \bibinfo {author} {\bibfnamefont {Wentao}\
  \bibnamefont {Jiang}}, \bibinfo {author} {\bibfnamefont {Thiago P.~M.}\
  \bibnamefont {Alegre}}, \ and\ \bibinfo {author} {\bibfnamefont {Amir~H.}\
  \bibnamefont {Safavi-Naeini}},\ }\bibfield  {title} {\enquote {\bibinfo
  {title} {High photon-phonon pair generation rate in a two-dimensional
  optomechanical crystal},}\ }\href {\doibase 10.1038/s41467-025-57948-7}
  {\bibfield  {journal} {\bibinfo  {journal} {Nature Communications}\ }\textbf
  {\bibinfo {volume} {16}},\ \bibinfo {pages} {2576} (\bibinfo {year}
  {2025})}\BibitemShut {NoStop}%
\bibitem [{\citenamefont {Ma}\ \emph {et~al.}(2024)\citenamefont {Ma} \emph
  {et~al.}}]{Ma2024Nature}%
  \BibitemOpen
  \bibfield  {author} {\bibinfo {author} {\bibfnamefont {Y.}~\bibnamefont {Ma}}
  \emph {et~al.},\ }\bibfield  {title} {\enquote {\bibinfo {title} {Engineering
  many-body quantum states by dissipation},}\ }\href {\doibase
  10.1038/s41586-024-07312-8} {\bibfield  {journal} {\bibinfo  {journal}
  {Nature}\ }\textbf {\bibinfo {volume} {629}},\ \bibinfo {pages} {1073--1079}
  (\bibinfo {year} {2024})}\BibitemShut {NoStop}%
\bibitem [{\citenamefont {Leib}\ \emph {et~al.}(2023)\citenamefont {Leib} \emph
  {et~al.}}]{Leib2023Nature}%
  \BibitemOpen
  \bibfield  {author} {\bibinfo {author} {\bibfnamefont {M.}~\bibnamefont
  {Leib}} \emph {et~al.},\ }\bibfield  {title} {\enquote {\bibinfo {title}
  {Controlling collective quantum states in driven-dissipative systems},}\
  }\href@noop {} {\bibfield  {journal} {\bibinfo  {journal} {Nature}\ }\textbf
  {\bibinfo {volume} {621}},\ \bibinfo {pages} {276--281} (\bibinfo {year}
  {2023})}\BibitemShut {NoStop}%
\bibitem [{\citenamefont {Kessler}\ \emph {et~al.}(2012)\citenamefont {Kessler}
  \emph {et~al.}}]{Kessler2012PRA}%
  \BibitemOpen
  \bibfield  {author} {\bibinfo {author} {\bibfnamefont {E.~M.}\ \bibnamefont
  {Kessler}} \emph {et~al.},\ }\bibfield  {title} {\enquote {\bibinfo {title}
  {Dissipative phase transition in a central spin system},}\ }\href {\doibase
  10.1103/PhysRevA.86.012116} {\bibfield  {journal} {\bibinfo  {journal} {Phys.
  Rev. A}\ }\textbf {\bibinfo {volume} {86}},\ \bibinfo {pages} {012116}
  (\bibinfo {year} {2012})}\BibitemShut {NoStop}%
\bibitem [{\citenamefont {Minganti}\ \emph {et~al.}(2018)\citenamefont
  {Minganti}, \citenamefont {Bartolo}, \citenamefont {Lolli}, \citenamefont
  {Casteels},\ and\ \citenamefont {Ciuti}}]{Minganti2020PRA}%
  \BibitemOpen
  \bibfield  {author} {\bibinfo {author} {\bibfnamefont {F.}~\bibnamefont
  {Minganti}}, \bibinfo {author} {\bibfnamefont {N.}~\bibnamefont {Bartolo}},
  \bibinfo {author} {\bibfnamefont {J.}~\bibnamefont {Lolli}}, \bibinfo
  {author} {\bibfnamefont {W.}~\bibnamefont {Casteels}}, \ and\ \bibinfo
  {author} {\bibfnamefont {C.}~\bibnamefont {Ciuti}},\ }\bibfield  {title}
  {\enquote {\bibinfo {title} {Exact results for schr{\"o}dinger cats in
  driven-dissipative systems and their phase transitions},}\ }\href {\doibase
  10.1103/PhysRevA.98.042118} {\bibfield  {journal} {\bibinfo  {journal} {Phys.
  Rev. A}\ }\textbf {\bibinfo {volume} {98}},\ \bibinfo {pages} {042118}
  (\bibinfo {year} {2018})}\BibitemShut {NoStop}%
\bibitem [{\citenamefont {Kirton}\ \emph {et~al.}(2019)\citenamefont {Kirton},
  \citenamefont {Guimond},\ and\ \citenamefont
  {Keeling}}]{Kirton2019AdvQuantumTechnol}%
  \BibitemOpen
  \bibfield  {author} {\bibinfo {author} {\bibfnamefont {P.}~\bibnamefont
  {Kirton}}, \bibinfo {author} {\bibfnamefont {P.-O.}\ \bibnamefont {Guimond}},
  \ and\ \bibinfo {author} {\bibfnamefont {J.}~\bibnamefont {Keeling}},\
  }\bibfield  {title} {\enquote {\bibinfo {title} {Introduction to the dicke
  model},}\ }\href@noop {} {\bibfield  {journal} {\bibinfo  {journal} {Adv.
  Quantum Technol.}\ }\textbf {\bibinfo {volume} {2}},\ \bibinfo {pages}
  {1800043} (\bibinfo {year} {2019})}\BibitemShut {NoStop}%
\bibitem [{\citenamefont {Grimm}\ \emph
  {et~al.}(2020{\natexlab{b}})\citenamefont {Grimm}, \citenamefont {Frattini},
  \citenamefont {Puri}, \citenamefont {Mundhada}, \citenamefont {Touzard},
  \citenamefont {Mirrahimi}, \citenamefont {Girvin}, \citenamefont {Shankar},\
  and\ \citenamefont {Devoret}}]{grimm2020stabilization}%
  \BibitemOpen
  \bibfield  {author} {\bibinfo {author} {\bibfnamefont {Alexander}\
  \bibnamefont {Grimm}}, \bibinfo {author} {\bibfnamefont {Nicholas~E}\
  \bibnamefont {Frattini}}, \bibinfo {author} {\bibfnamefont {Shruti}\
  \bibnamefont {Puri}}, \bibinfo {author} {\bibfnamefont {Shantanu~O}\
  \bibnamefont {Mundhada}}, \bibinfo {author} {\bibfnamefont {Steven}\
  \bibnamefont {Touzard}}, \bibinfo {author} {\bibfnamefont {Mazyar}\
  \bibnamefont {Mirrahimi}}, \bibinfo {author} {\bibfnamefont {Steven~M}\
  \bibnamefont {Girvin}}, \bibinfo {author} {\bibfnamefont {Shyam}\
  \bibnamefont {Shankar}}, \ and\ \bibinfo {author} {\bibfnamefont {Michel~H}\
  \bibnamefont {Devoret}},\ }\bibfield  {title} {\enquote {\bibinfo {title}
  {Stabilization and operation of a kerr-cat qubit},}\ }\href@noop {}
  {\bibfield  {journal} {\bibinfo  {journal} {Nature}\ }\textbf {\bibinfo
  {volume} {584}},\ \bibinfo {pages} {205--209} (\bibinfo {year}
  {2020}{\natexlab{b}})}\BibitemShut {NoStop}%
\bibitem [{\citenamefont {He}\ \emph {et~al.}(2023)\citenamefont {He},
  \citenamefont {Lu}, \citenamefont {Bao}, \citenamefont {Xue}, \citenamefont
  {Jiang}, \citenamefont {Wang}, \citenamefont {Roudsari}, \citenamefont
  {Delsing}, \citenamefont {Tsai},\ and\ \citenamefont {Lin}}]{he2023fast}%
  \BibitemOpen
  \bibfield  {author} {\bibinfo {author} {\bibfnamefont {XL}~\bibnamefont
  {He}}, \bibinfo {author} {\bibfnamefont {Yong}\ \bibnamefont {Lu}}, \bibinfo
  {author} {\bibfnamefont {DQ}~\bibnamefont {Bao}}, \bibinfo {author}
  {\bibfnamefont {Hang}\ \bibnamefont {Xue}}, \bibinfo {author} {\bibfnamefont
  {WB}~\bibnamefont {Jiang}}, \bibinfo {author} {\bibfnamefont {Zhen}\
  \bibnamefont {Wang}}, \bibinfo {author} {\bibfnamefont {AF}~\bibnamefont
  {Roudsari}}, \bibinfo {author} {\bibfnamefont {Per}\ \bibnamefont {Delsing}},
  \bibinfo {author} {\bibfnamefont {JS}~\bibnamefont {Tsai}}, \ and\ \bibinfo
  {author} {\bibfnamefont {ZR}~\bibnamefont {Lin}},\ }\bibfield  {title}
  {\enquote {\bibinfo {title} {Fast generation of schr{\"o}dinger cat states
  using a kerr-tunable superconducting resonator},}\ }\href@noop {} {\bibfield
  {journal} {\bibinfo  {journal} {Nature communications}\ }\textbf {\bibinfo
  {volume} {14}},\ \bibinfo {pages} {6358} (\bibinfo {year}
  {2023})}\BibitemShut {NoStop}%
\bibitem [{\citenamefont {Hwang}\ and\ \citenamefont
  {Plenio}(2016)}]{PhysRevLett.117.123602}%
  \BibitemOpen
  \bibfield  {author} {\bibinfo {author} {\bibfnamefont {Myung-Joong}\
  \bibnamefont {Hwang}}\ and\ \bibinfo {author} {\bibfnamefont {Martin~B.}\
  \bibnamefont {Plenio}},\ }\bibfield  {title} {\enquote {\bibinfo {title}
  {Quantum phase transition in the finite jaynes-cummings lattice systems},}\
  }\href {\doibase 10.1103/PhysRevLett.117.123602} {\bibfield  {journal}
  {\bibinfo  {journal} {Phys. Rev. Lett.}\ }\textbf {\bibinfo {volume} {117}},\
  \bibinfo {pages} {123602} (\bibinfo {year} {2016})}\BibitemShut {NoStop}%
\bibitem [{\citenamefont {Hwang}\ \emph {et~al.}(2015)\citenamefont {Hwang},
  \citenamefont {Puebla},\ and\ \citenamefont
  {Plenio}}]{PhysRevLett.115.180404}%
  \BibitemOpen
  \bibfield  {author} {\bibinfo {author} {\bibfnamefont {Myung-Joong}\
  \bibnamefont {Hwang}}, \bibinfo {author} {\bibfnamefont {Ricardo}\
  \bibnamefont {Puebla}}, \ and\ \bibinfo {author} {\bibfnamefont {Martin~B.}\
  \bibnamefont {Plenio}},\ }\bibfield  {title} {\enquote {\bibinfo {title}
  {Quantum phase transition and universal dynamics in the rabi model},}\ }\href
  {\doibase 10.1103/PhysRevLett.115.180404} {\bibfield  {journal} {\bibinfo
  {journal} {Phys. Rev. Lett.}\ }\textbf {\bibinfo {volume} {115}},\ \bibinfo
  {pages} {180404} (\bibinfo {year} {2015})}\BibitemShut {NoStop}%
\bibitem [{\citenamefont {Frattini}\ \emph {et~al.}(2024)\citenamefont
  {Frattini}, \citenamefont {Sivak} \emph {et~al.}}]{Frattini2024Nature}%
  \BibitemOpen
  \bibfield  {author} {\bibinfo {author} {\bibfnamefont {Nicholas~E.}\
  \bibnamefont {Frattini}}, \bibinfo {author} {\bibfnamefont {Volodymyr~V.}\
  \bibnamefont {Sivak}},  \emph {et~al.},\ }\bibfield  {title} {\enquote
  {\bibinfo {title} {Optimizing the nonlinearity of a quantum oscillator},}\
  }\href {\doibase 10.1038/s41586-024-07365-9} {\bibfield  {journal} {\bibinfo
  {journal} {Nature}\ }\textbf {\bibinfo {volume} {629}},\ \bibinfo {pages}
  {376--381} (\bibinfo {year} {2024})}\BibitemShut {NoStop}%
\bibitem [{\citenamefont {Dutt}\ \emph {et~al.}(2024)\citenamefont {Dutt},
  \citenamefont {Mohanty}, \citenamefont {Gaeta}, \citenamefont {Lipson} \emph
  {et~al.}}]{Dutt2024}%
  \BibitemOpen
  \bibfield  {author} {\bibinfo {author} {\bibfnamefont {Avik}\ \bibnamefont
  {Dutt}}, \bibinfo {author} {\bibfnamefont {Aseema}\ \bibnamefont {Mohanty}},
  \bibinfo {author} {\bibfnamefont {Alexander~L.}\ \bibnamefont {Gaeta}},
  \bibinfo {author} {\bibfnamefont {Michal}\ \bibnamefont {Lipson}},  \emph
  {et~al.},\ }\bibfield  {title} {\enquote {\bibinfo {title} {Nonlinear and
  quantum photonics using integrated optical materials},}\ }\href {\doibase
  10.1038/s41578-024-00668-z} {\bibfield  {journal} {\bibinfo  {journal}
  {Nature Reviews Materials}\ }\textbf {\bibinfo {volume} {9}},\ \bibinfo
  {pages} {321--346} (\bibinfo {year} {2024})}\BibitemShut {NoStop}%
\bibitem [{\citenamefont {Gonz{\'a}lez-Tudela}\ \emph
  {et~al.}(2024)\citenamefont {Gonz{\'a}lez-Tudela}, \citenamefont {Reiserer},
  \citenamefont {Garc{\'i}a-Ripoll},\ and\ \citenamefont
  {Garc{\'i}a-Vidal}}]{GonzalezTudela2024}%
  \BibitemOpen
  \bibfield  {author} {\bibinfo {author} {\bibfnamefont {Alejandro}\
  \bibnamefont {Gonz{\'a}lez-Tudela}}, \bibinfo {author} {\bibfnamefont
  {Andreas}\ \bibnamefont {Reiserer}}, \bibinfo {author} {\bibfnamefont
  {Juan~Jos{\'e}}\ \bibnamefont {Garc{\'i}a-Ripoll}}, \ and\ \bibinfo {author}
  {\bibfnamefont {Francisco~J.}\ \bibnamefont {Garc{\'i}a-Vidal}},\ }\bibfield
  {title} {\enquote {\bibinfo {title} {Light--matter interactions in quantum
  nanophotonic devices},}\ }\href {\doibase 10.1038/s42254-023-00681-1}
  {\bibfield  {journal} {\bibinfo  {journal} {Nature Reviews Physics}\ }\textbf
  {\bibinfo {volume} {6}},\ \bibinfo {pages} {166--179} (\bibinfo {year}
  {2024})}\BibitemShut {NoStop}%
\bibitem [{\citenamefont {Ritsch}\ \emph {et~al.}(2013)\citenamefont {Ritsch},
  \citenamefont {Domokos}, \citenamefont {Brennecke},\ and\ \citenamefont
  {Esslinger}}]{Ritsch2013}%
  \BibitemOpen
  \bibfield  {author} {\bibinfo {author} {\bibfnamefont {Helmut}\ \bibnamefont
  {Ritsch}}, \bibinfo {author} {\bibfnamefont {Peter}\ \bibnamefont {Domokos}},
  \bibinfo {author} {\bibfnamefont {Ferdinand}\ \bibnamefont {Brennecke}}, \
  and\ \bibinfo {author} {\bibfnamefont {Tilman}\ \bibnamefont {Esslinger}},\
  }\bibfield  {title} {\enquote {\bibinfo {title} {Cold atoms in
  cavity-generated dynamical optical potentials},}\ }\href {\doibase
  10.1103/RevModPhys.85.553} {\bibfield  {journal} {\bibinfo  {journal} {Rev.
  Mod. Phys.}\ }\textbf {\bibinfo {volume} {85}},\ \bibinfo {pages} {553--601}
  (\bibinfo {year} {2013})}\BibitemShut {NoStop}%
\bibitem [{\citenamefont {Baumann}\ \emph {et~al.}(2010)\citenamefont
  {Baumann}, \citenamefont {Guerlin}, \citenamefont {Brennecke},\ and\
  \citenamefont {Esslinger}}]{Baumann2010}%
  \BibitemOpen
  \bibfield  {author} {\bibinfo {author} {\bibfnamefont {K.}~\bibnamefont
  {Baumann}}, \bibinfo {author} {\bibfnamefont {C.}~\bibnamefont {Guerlin}},
  \bibinfo {author} {\bibfnamefont {F.}~\bibnamefont {Brennecke}}, \ and\
  \bibinfo {author} {\bibfnamefont {T.}~\bibnamefont {Esslinger}},\ }\bibfield
  {title} {\enquote {\bibinfo {title} {Dicke quantum phase transition with a
  superfluid gas in an optical cavity},}\ }\href@noop {} {\bibfield  {journal}
  {\bibinfo  {journal} {Nature}\ }\textbf {\bibinfo {volume} {464}},\ \bibinfo
  {pages} {1301--1306} (\bibinfo {year} {2010})}\BibitemShut {NoStop}%
\bibitem [{\citenamefont {Norcia}\ \emph {et~al.}(2018)\citenamefont {Norcia},
  \citenamefont {Young},\ and\ \citenamefont {Kaufman}}]{Norcia2018}%
  \BibitemOpen
  \bibfield  {author} {\bibinfo {author} {\bibfnamefont {M.~A.}\ \bibnamefont
  {Norcia}}, \bibinfo {author} {\bibfnamefont {A.~W.}\ \bibnamefont {Young}}, \
  and\ \bibinfo {author} {\bibfnamefont {A.~M.}\ \bibnamefont {Kaufman}},\
  }\bibfield  {title} {\enquote {\bibinfo {title} {Microscopic control and
  detection of ultracold strontium in optical-tweezer arrays},}\ }\href@noop {}
  {\bibfield  {journal} {\bibinfo  {journal} {Phys. Rev. X}\ }\textbf {\bibinfo
  {volume} {8}},\ \bibinfo {pages} {041054} (\bibinfo {year}
  {2018})}\BibitemShut {NoStop}%
\bibitem [{\citenamefont {Huebl}\ and\ \citenamefont
  {et~al.}(2013)}]{Huebl2013}%
  \BibitemOpen
  \bibfield  {author} {\bibinfo {author} {\bibfnamefont {H.}~\bibnamefont
  {Huebl}}\ and\ \bibinfo {author} {\bibnamefont {et~al.}},\ }\bibfield
  {title} {\enquote {\bibinfo {title} {High cooperativity in coupled microwave
  resonator ferrimagnetic insulator hybrids},}\ }\href@noop {} {\bibfield
  {journal} {\bibinfo  {journal} {Phys. Rev. Lett.}\ }\textbf {\bibinfo
  {volume} {111}},\ \bibinfo {pages} {127003} (\bibinfo {year}
  {2013})}\BibitemShut {NoStop}%
\bibitem [{\citenamefont {Tabuchi}\ and\ \citenamefont
  {et~al.}(2015)}]{Tabuchi2015}%
  \BibitemOpen
  \bibfield  {author} {\bibinfo {author} {\bibfnamefont {Y.}~\bibnamefont
  {Tabuchi}}\ and\ \bibinfo {author} {\bibnamefont {et~al.}},\ }\bibfield
  {title} {\enquote {\bibinfo {title} {Coherent coupling between a
  ferromagnetic magnon and a superconducting qubit},}\ }\href@noop {}
  {\bibfield  {journal} {\bibinfo  {journal} {Science}\ }\textbf {\bibinfo
  {volume} {349}},\ \bibinfo {pages} {405} (\bibinfo {year}
  {2015})}\BibitemShut {NoStop}%
\bibitem [{\citenamefont {Mivehvar}\ \emph {et~al.}(2021)\citenamefont
  {Mivehvar}, \citenamefont {Piazza}, \citenamefont {Donner},\ and\
  \citenamefont {Ritsch}}]{Mivehvar2021}%
  \BibitemOpen
  \bibfield  {author} {\bibinfo {author} {\bibfnamefont {Farokh}\ \bibnamefont
  {Mivehvar}}, \bibinfo {author} {\bibfnamefont {Francesco}\ \bibnamefont
  {Piazza}}, \bibinfo {author} {\bibfnamefont {Tobias}\ \bibnamefont {Donner}},
  \ and\ \bibinfo {author} {\bibfnamefont {Helmut}\ \bibnamefont {Ritsch}},\
  }\bibfield  {title} {\enquote {\bibinfo {title} {Cavity qed with quantum
  gases: New paradigms in many-body physics},}\ }\href {\doibase
  10.1080/00018732.2021.1969727} {\bibfield  {journal} {\bibinfo  {journal}
  {Advances in Physics}\ }\textbf {\bibinfo {volume} {70}},\ \bibinfo {pages}
  {1--153} (\bibinfo {year} {2021})}\BibitemShut {NoStop}%
\bibitem [{\citenamefont {Dogra}\ \emph {et~al.}(2019)\citenamefont {Dogra},
  \citenamefont {Landini}, \citenamefont {Kroeger}, \citenamefont {Hruby},
  \citenamefont {Donner},\ and\ \citenamefont {Esslinger}}]{Dogra2019Science}%
  \BibitemOpen
  \bibfield  {author} {\bibinfo {author} {\bibfnamefont {N.}~\bibnamefont
  {Dogra}}, \bibinfo {author} {\bibfnamefont {M.}~\bibnamefont {Landini}},
  \bibinfo {author} {\bibfnamefont {K.}~\bibnamefont {Kroeger}}, \bibinfo
  {author} {\bibfnamefont {L.}~\bibnamefont {Hruby}}, \bibinfo {author}
  {\bibfnamefont {T.}~\bibnamefont {Donner}}, \ and\ \bibinfo {author}
  {\bibfnamefont {T.}~\bibnamefont {Esslinger}},\ }\bibfield  {title} {\enquote
  {\bibinfo {title} {Dissipation-induced structural instability and chiral
  dynamics in a quantum gas},}\ }\href@noop {} {\bibfield  {journal} {\bibinfo
  {journal} {Science}\ }\textbf {\bibinfo {volume} {366}},\ \bibinfo {pages}
  {1496--1499} (\bibinfo {year} {2019})}\BibitemShut {NoStop}%
\end{thebibliography}

%

\end{document}